\documentclass[journal]{IEEEtran}

\ifCLASSINFOpdf
\else

\fi

\usepackage{amsmath}
\usepackage{esint,color}
\usepackage{subfigure}
\usepackage{gensymb}
\usepackage{cite}
\usepackage{graphicx}

\begin{document}

\bstctlcite{IEEEexample:BSTcontrol}

%
\title{Design of a Single-Element Dynamic Antenna for Secure Wireless Applications}
\author{Amer Abu Arisheh,~\IEEEmembership{Graduate Student Member,~IEEE}, Jason M. Merlo,~\IEEEmembership{Graduate Student Member,~IEEE},\\and Jeffrey A. Nanzer,~\IEEEmembership{Senior Member,~IEEE}
\thanks{This material is based in part upon work supported by the National Science Foundation under grant number 2028736. \textit{(Corresponding author: Jeffrey A. Nanzer)}}
\thanks{The authors are with the Department of Electrical and Computer Engineering, Michigan State University, East Lansing, MI 48824 USA (email: \{abuarish, merlojas, nanzer\}@msu.edu).}
} %

\markboth{IEEE}%
{Shell \MakeLowercase{\textit{et al.}}: Bare Demo of IEEEtran.cls for IEEE Journals}


%



\maketitle


\begin{abstract}


We introduce a new technique for secure wireless applications using a single dynamic antenna. The dynamic antenna supports a constantly changing current distribution that generates a radiation pattern that is static in a desired direction and dynamic elsewhere, thereby imparting additional modulation on the signal and obscuring information transmitted or received outside of the secure spatial region. Dynamic currents are supported by a single feed that is switched between separate ports on a single antenna, generating two different radiation patterns. We introduce the theoretical concept by exploring an ideal complex dynamic radiation pattern that remains static in a narrow desired direction and is dynamic elsewhere. The impact on the transmission of information is analyzed, showing that the secure region narrows as the modulation order increases, and design constraints on the spatial width of the secure region as a function of modulation format are determined. We design and analyze a 2.3 GHz two-state dynamic dipole antenna and experimentally demonstrate secure wireless transmission. 
We show the ability to steer the secure region experimentally, and to maintain high throughput in the secure region while obscuring the information elsewhere. Our approach introduces a novel single-element technique for secure wireless applications that can be used independently from the rest of the wireless system, essentially operating as a ``black box'' for an additional layer of security.


\end{abstract}


%
\IEEEpeerreviewmaketitle

\section{Introduction}
\label{Introduction}

Security is rapidly becoming a central aspect of wireless system design. With the increase in wireless communications devices as well as wireless sensing systems, all of which are becoming increasingly connected via wireless networks, the potential for vulnerabilities, both from inadvertent and malicious means, is commensurately increasing. 
Vulnerabilities in wireless networks are an active area of research~\cite{7467419}, however research on the security of emerging sensor systems such as automotive radar have also demonstrated the potential physical layer vulnerabilities of such systems~\cite{miller2018securing,8690734}. 
While security approaches for wireless systems have traditionally focused on digital security, there is significant interest in and potential for embedding security into the physical layer of wireless systems to serve as a first line of defense against unintended or malicious actions. Security at the physical layer reduces the burden on digital security protocols, thereby supporting better efficiency and throughput, and can potentially be implemented as a black box protocol that is transparent to the rest of the wireless system. 

Various physical layer approaches to wireless security have been explored (see, e.g.,~\cite{5751298}), including radio-frequency (RF) fingerprinting, signal coding, and multiple-input multiple-output (MIMO) architectures. However, the approaches that are closest to the physical wireless transmission and have the potential for little to no coordination with the rest of the system use the antenna system itself. 
Directional antennas or antenna arrays steer higher gain signal towards intended directions, with lower gain elsewhere, leading to reduced detectability and information recovery outside the mainbeam~\cite{534916}. The antenna pattern is generally static and thus the same signal is broadcast to all directions; therefore, an eavesdropper that is close to the antenna or has a more sensitive receiver can still potentially recover the information. 
Dynamic antenna patterns have been implemented to overcome this problem using additional modulation imparted at the antenna elements within the array structure, creating an additional directional modulation onto the transmitted (or received) signal at angles outside the mainbeam, thereby obfuscating the information and making it more difficult if not impossible to demodulate~\cite{daly2009directional,Daly3,Daly4,ding2013vector}. Such arrays have been implemented by controlling the excitation weights of each array antenna element~\cite{daly2010demonstration}, by reconfiguring the antenna radiating elements~\cite{babakhani2008transmitter}, or by using dynamic motion in distributed antenna arrays~\cite{9665259}.
But while these approaches are feasible for array applications, many wireless systems use compact antennas, often single-elements, in which physical layer security is more difficult to impart. Some works have investigated the use of multiple antenna elements constructed within a compact space (e.g.,~\cite{2017,2021}), however these nonetheless require multiple signal feeds in order to impart directional modulation, increasing the complexity of the wireless system and preventing the physical layer approach from being transparent to the rest of the system.

\begin{figure*}[t!]
\centering
\includegraphics[width=0.8\textwidth]{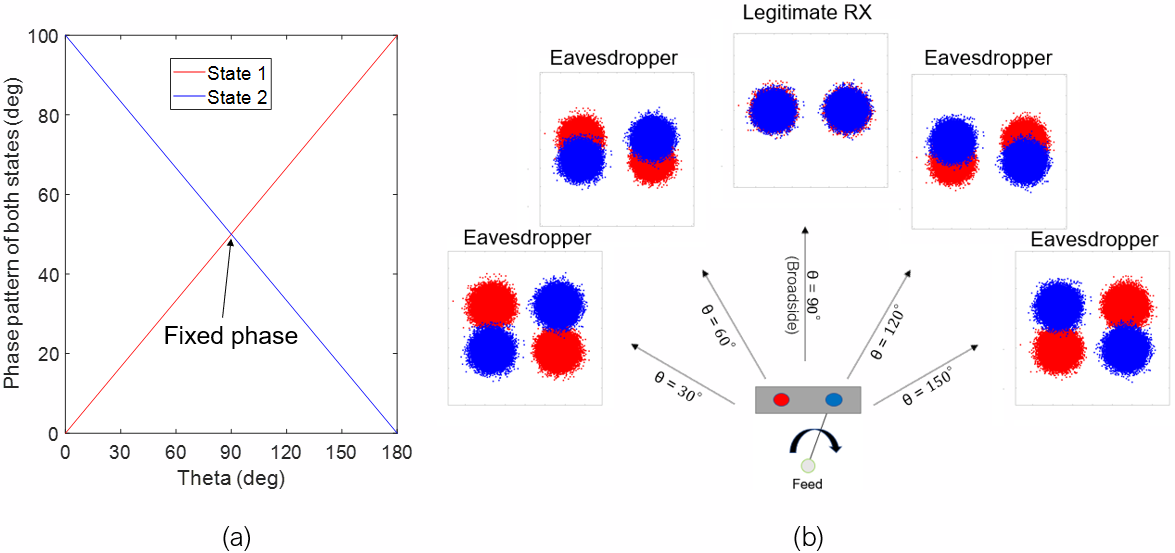}
\caption{Phase dynamics of a hypothesized antenna for directional modulation. (a) Phase patterns generated by a notional antenna by connecting the feeding port to the red feed point (State 1) and by connecting the feeding port to the blue feed point (State 2); isotropic amplitude patterns are assumed. (b) Constellations obtained by modulating a pseudo random bit sequence (PRBS) by the phase patterns with 12 dB SNR. The signals are standard in the direction of the legitimate receiver, but distorted elsewhere.}
\label{Phase_Dynamics_BPSK}
\end{figure*}

In this paper, we present a new approach to physical layer wireless security based on a novel dynamic single-element antenna. Through switching of the current path on the antenna structure, the current distribution on the element is dynamically modified, generating a dynamic radiation pattern that remains static in a desired direction and imparts additional amplitude and phase modulation in directions away from the secure region. In contrast to reconfigurable antennas, which typically cycle through various states to find a desired radiation pattern, our dynamic antenna design is characterized by constant change in the radiation pattern, thereby creating additional time-varying modulation on the signal outside the secure region. 
The approach is simple and requires no computational analysis of the antenna states, similar to other synthesis-free directional modulation approaches~\cite{valliappan2013antenna,zhu2013directional,ding2016synthesis}.
We introduced the basics of the concept in~\cite{Our_URSI,Our_APS}; here we provide extensive detail on the underlying theory behind dynamic antennas and their use in directional modulation for various modulation formats and describe the ability to steer the secure region. We describe in detail the impact of dynamic amplitude and phase patterns. We present the design and the first experimental implementation of a 2.3 GHz dynamic dipole antenna, and experimentally demonstrate the use of the dynamic antenna for secure transmission of data in a wireless communications system. 
Our experiments show the ability to retain high throughput in the desired direction while mitigating throughput outside the secure region using dynamics at the same speed as the data. However, the presented approach need not dynamically switch at the same speed as the information, as lower or higher rates could be used if the application allows. Importantly, this means that the dynamic antenna system can operate effectively independent of the underlying wireless system, such that it can be used as an external add-on to existing wireless systems.
Furthermore, while we show the use of communications systems, the approach also applies to secure wireless sensing (on both transmit and receive). 

The rest of this paper is outlined as follows. In Section~\ref{Theory} we present the foundational concept in detail, describing the effects of an ideal dynamic radiation pattern on information throughput as a function of angle. A numerical assessment is provided in Section~\ref{Numerical}. In Section \ref{Design}, a $3\lambda/2$ printed dipole antenna is designed based on the theoretical concepts from Section~\ref{Theory}. In Section~\ref{RP_Measurements}, the dynamic antenna is fabricated and measured, and Section~\ref{BER_Measurements} presents comprehensive measurements of the antenna in a wireless communication system. In Section~\ref{Steering_Section}, we demonstrate how the secure region can be steered. Finally, Section~\ref{Conclusion} concludes this work.

\section{Theory of Directional Modulation by a Dynamic Single Antenna}
\label{Theory}

\begin{figure}[t!]
\centering
\includegraphics[width=0.48\textwidth]{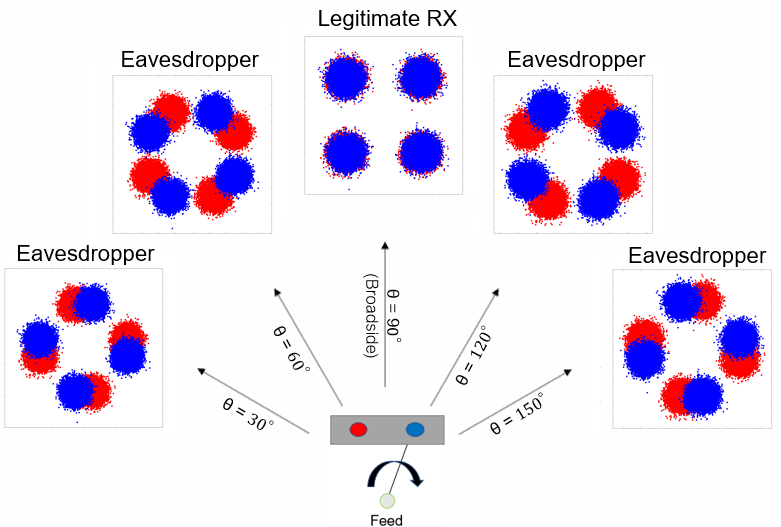}
\caption{Phase dynamics of a hypothesized antenna achieving directional modulation for 4-QAM.}
\label{Phase_Dynamics_QAM16}
\end{figure}

\begin{figure}[t!]
\centering
\includegraphics[width=0.48\textwidth]{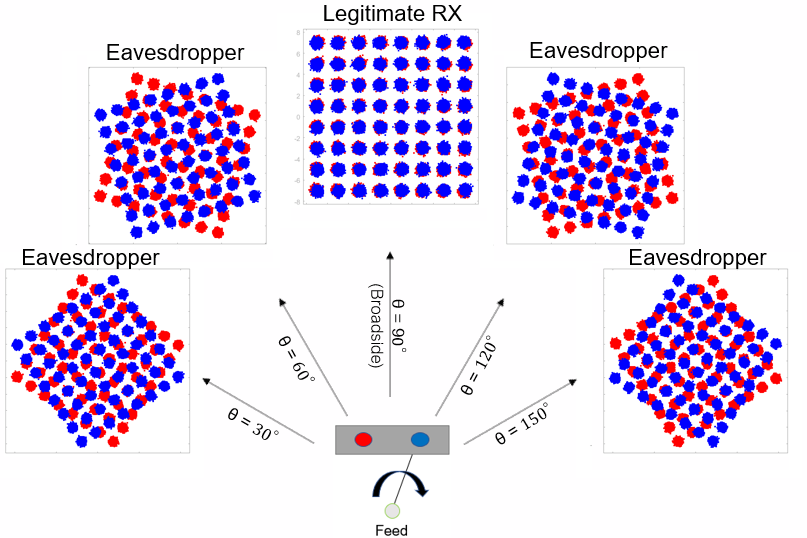}
\caption{Phase dynamics of a hypothesized antenna achieving directional modulation for 64-QAM.}
\label{Phase_Dynamics_QAM64}
\end{figure}

\begin{figure}[t!]
\centering
\includegraphics[width=0.35\textwidth]{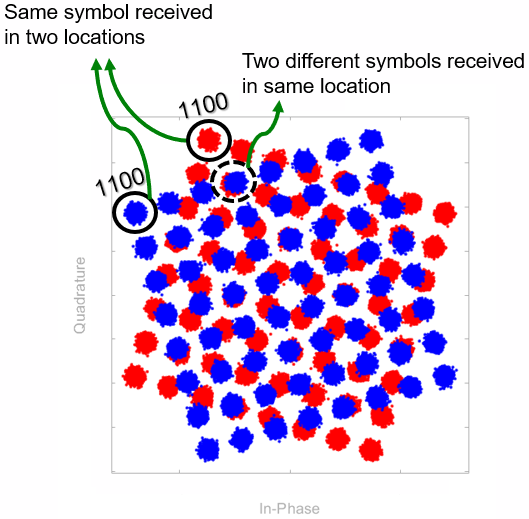}
\caption{Phase dynamics distort the constellation of symbols via rotation. The summation is clearly distorted by shifting symbols to different locations, but also sometimes with different symbols received in the same location.}
\label{Example}
\end{figure}

The radiated fields of any antenna are governed by the current flowing in that antenna. Thus, in general, feeding an antenna from two different feed points yields different radiation patterns versus angle. We start abstractly without confining ourselves to any particular antenna and consider a general single-port two-state switched antenna in Fig. \ref{Phase_Dynamics_BPSK}, which shows the two feed points colored in red and blue. This antenna has a single feeding port and a switch that connects it to either the red or blue feed point, creating two states of the antenna. Fig. \ref{Phase_Dynamics_BPSK}(a) shows notional phase patterns radiated when feeding the antenna through the red or blue feed point, respectively. We assume the amplitude to be static versus angle for both states. While the physical realizability of such an antenna is not considered, the theoretical radiation pattern serves as a construct to demonstrate the impact of dynamic phase patterns. Fig. \ref{Phase_Dynamics_BPSK}(b) shows the scatter diagram (constellation) of a demodulated binary phase shift keyed signal received be the two states at various angles. The constellations are obtained by modulating a pseudo random bit sequence (PRBS) by the phase pattern shown in Fig. \ref{Phase_Dynamics_BPSK} while assuming isotropic amplitude pattern with 12 dB SNR. It is assumed that the legitimate receiver is in the broadside direction while eavesdroppers exist elsewhere. The received constellation at the broadside $\theta = 90^{\degree}$ is the standard binary phase shift keying (BPSK) diagram and is equal for both states since the phase pattern is equal at broadside. This implies that the legitimate receiver will not be affected by switching between the two states of the antenna. However, away from broadside the phase patterns of the two states impart additional phases, yielding rotation between the two constellations. This implies that the eavesdroppers will receive non-standard constellations. For example, at $\theta = 30^{\degree}$, we notice that the eavesdropper is receiving a constellation that looks similar to a quadrature-phase shift keyed (QPSK) signal. 
The impact of the phase rotation on the data is dependent on the modulation format. Increasing the transmitted modulation order to 4-level quadrature amplitude modulation (4-QAM) as shown in Fig. \ref{Phase_Dynamics_QAM16} or 64-QAM as shown in Fig. \ref{Phase_Dynamics_QAM64}, the constellations away from broadside are more affected by the phase dynamics since there is less phase difference between adjacent symbols. 
Since the eavesdropper sees the constellation diagram as the aggregate of the diagrams of the two states, it may be difficult or impossible to demodulate the underlying data. Furthermore, at some angles, the phase rotation may be such that separate symbols in each state land on the same point in the constellation (see Fig.~\ref{Example}), making demodulation more difficult. 

\begin{figure*}[t!]
\centering
\includegraphics[width=0.8\textwidth]{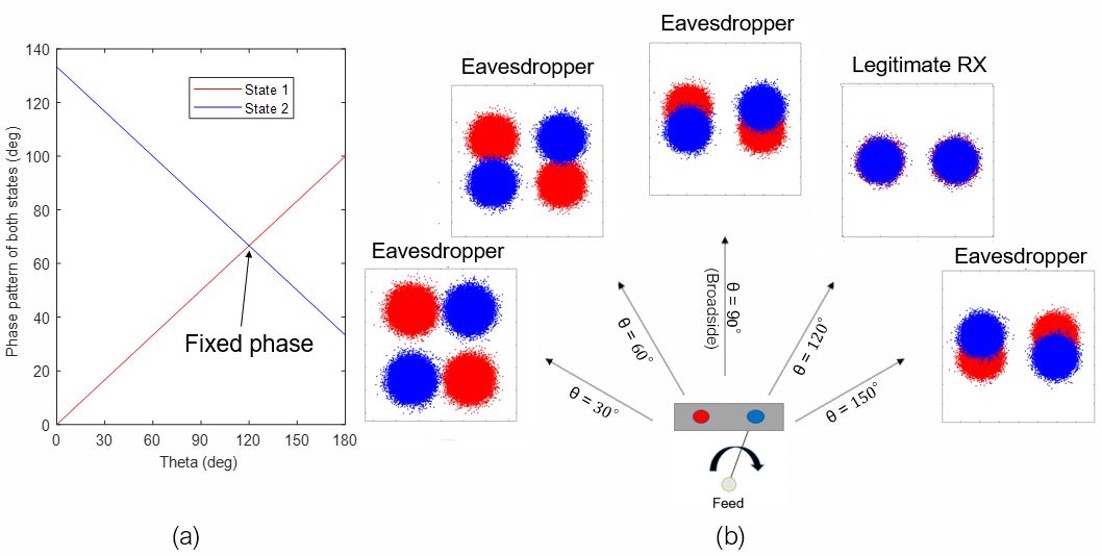}
\caption{Phase dynamics of the antenna after steering secure region to $\theta$ = 120$\degree$. (a) Phase patterns generated by adding a constant phase shift to port 2 (State 2) which equals the difference between the two phase patterns at $\theta$ = 120$\degree$; consequently, this yields a static phase at that angle. (b) Constellations obtained by modulating a PRBS by the phase patterns with 12 dB SNR.}
\label{steered}
\end{figure*}

Steering can be accomplished directly by weighting (in amplitude or phase) the signal prior to the feedpoint(s).
For example, to steer the secure region in Fig. \ref{Phase_Dynamics_BPSK} to $\theta = 120^{\degree}$, a phase shift to state 2 equal to the difference between the two states at $\theta = 120^{\degree}$ which equals 33.33$\degree$ is added, producing static fields in that desired direction as shown in Fig \ref{steered}. 

\begin{figure*}[t!]
\centering
\includegraphics[width=0.8\textwidth]{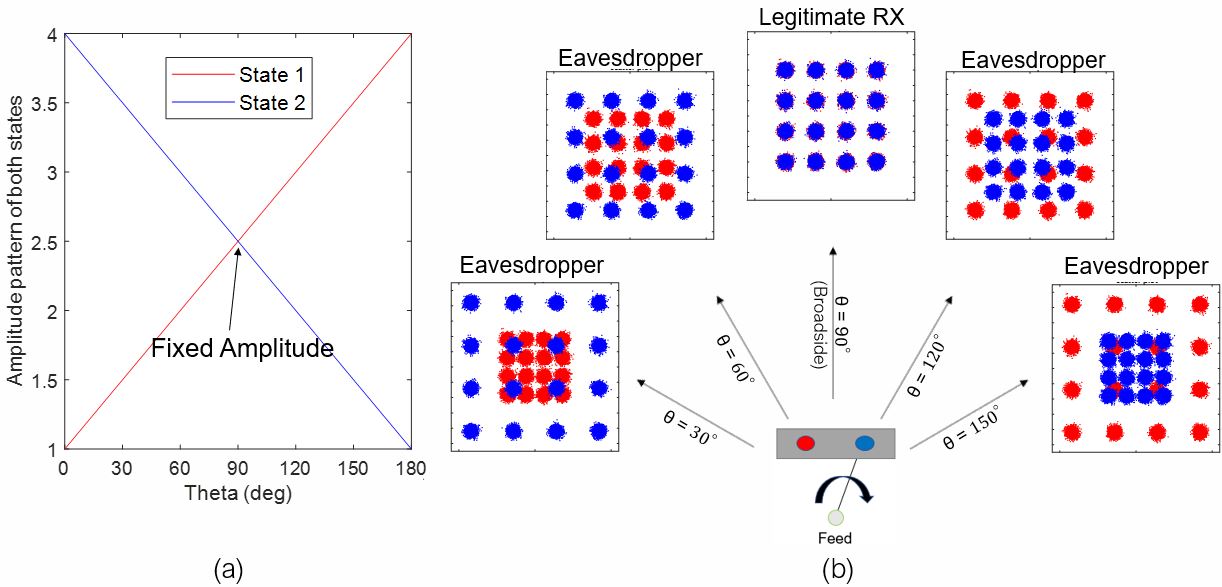}
\caption{Amplitude dynamics of a hypothesized antenna for directional modulation. (a) Amplitude patterns generated by a notional antenna by connecting the feeding port to the red feed point (State 1) and by connecting the feeding port to the blue feed point (State 2); isotropic phase patterns are assumed. (b) Constellations obtained by modulating a PRBS by the amplitude patterns with 12 dB SNR.}
\label{Amplitude_Dynamics_QAM16}
\end{figure*}

While the above analysis focused on phase dynamics, similar results can be obtained with amplitude dynamics.
The previous discussion of phase dynamics assumed an isotropic amplitude pattern. We now analyze the complementary problem by assuming that the radiated fields exhibit isotropic phase patterns, while amplitude patterns are assumed to be as shown in Fig. \ref{Amplitude_Dynamics_QAM16}(a). (Similar to the phase pattern analysis, we consider this theoretical radiation pattern as a construct to demonstrate the impact of dynamic amplitude patterns.) Consequently, Fig. \ref{Amplitude_Dynamics_QAM16}(b) shows the received constellations with 12 dB SNR at the broadside direction and 16-QAM modulation. The constellation in the broadside is standard due to the fixed amplitude, while away from broadside the constellation is distorted due to the amplitude difference between the two states. 

While some other works configure systems for a specific digital modulation
scheme, this technique works directly for any QPSK or QAM modulation order \cite{daly2009directional,daly2010demonstration,Daly3,Daly4}. Also, standard constellations can be obtained in our technique by making a perfect amplitude and phase calibration, in contrast to some other techniques. Moreover, the antenna can operate normally without exhibiting dynamics by simply fixing the antenna to operate to only one of the states. 

\section{Simulation of Information Transfer with a Dynamic Antenna}
\label{Numerical}

\begin{figure}
    \centering
     \subfigure[]
    {
        \includegraphics[width=0.48\textwidth]{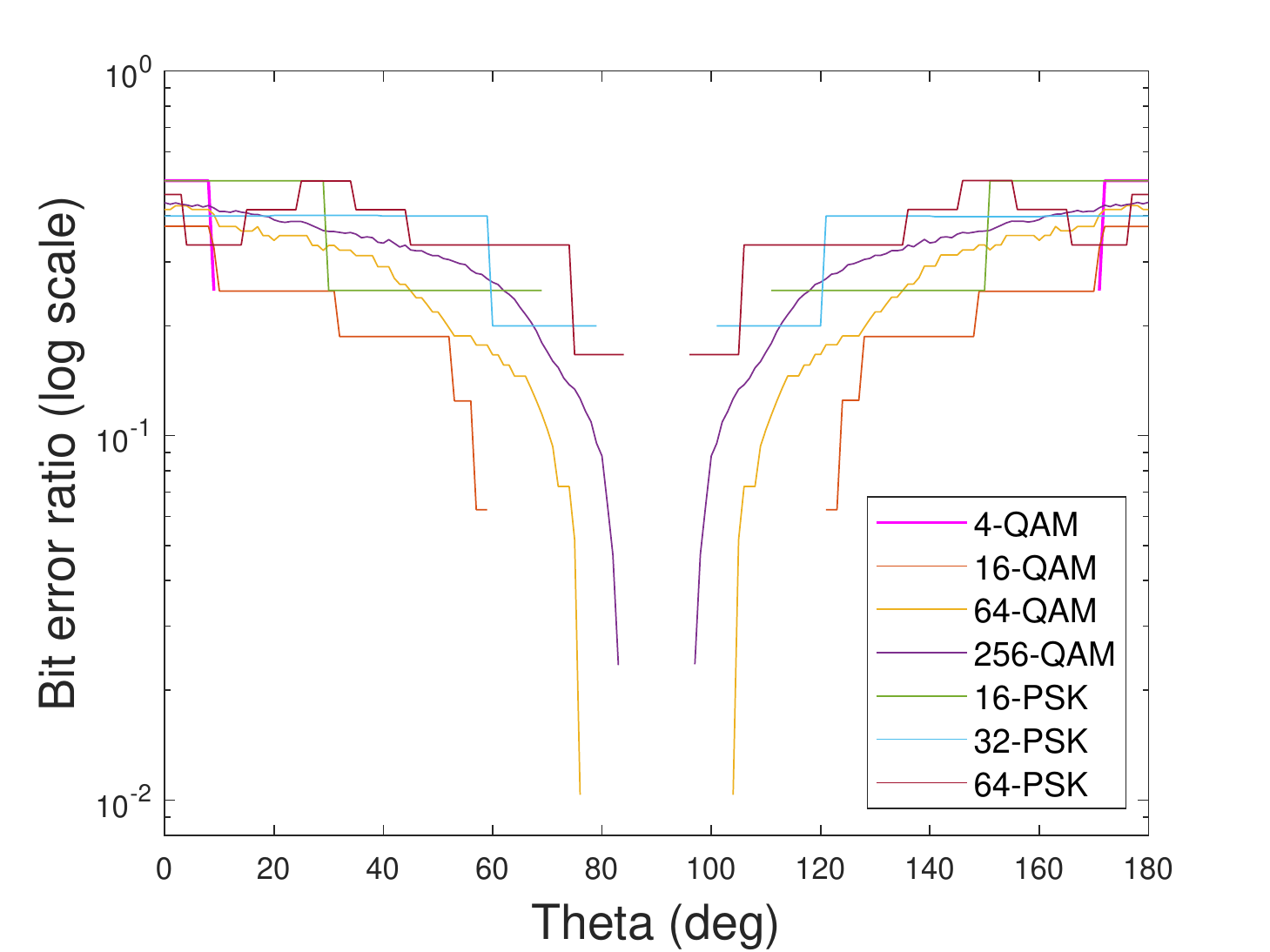}
       \label{PhaseBER}
    }
    \\
    \subfigure[The difference between the two states phase patterns with the thresholds lines below which the corresponding modulation schemes exhibit zero BER.]
    {
        \includegraphics[width=0.48\textwidth]{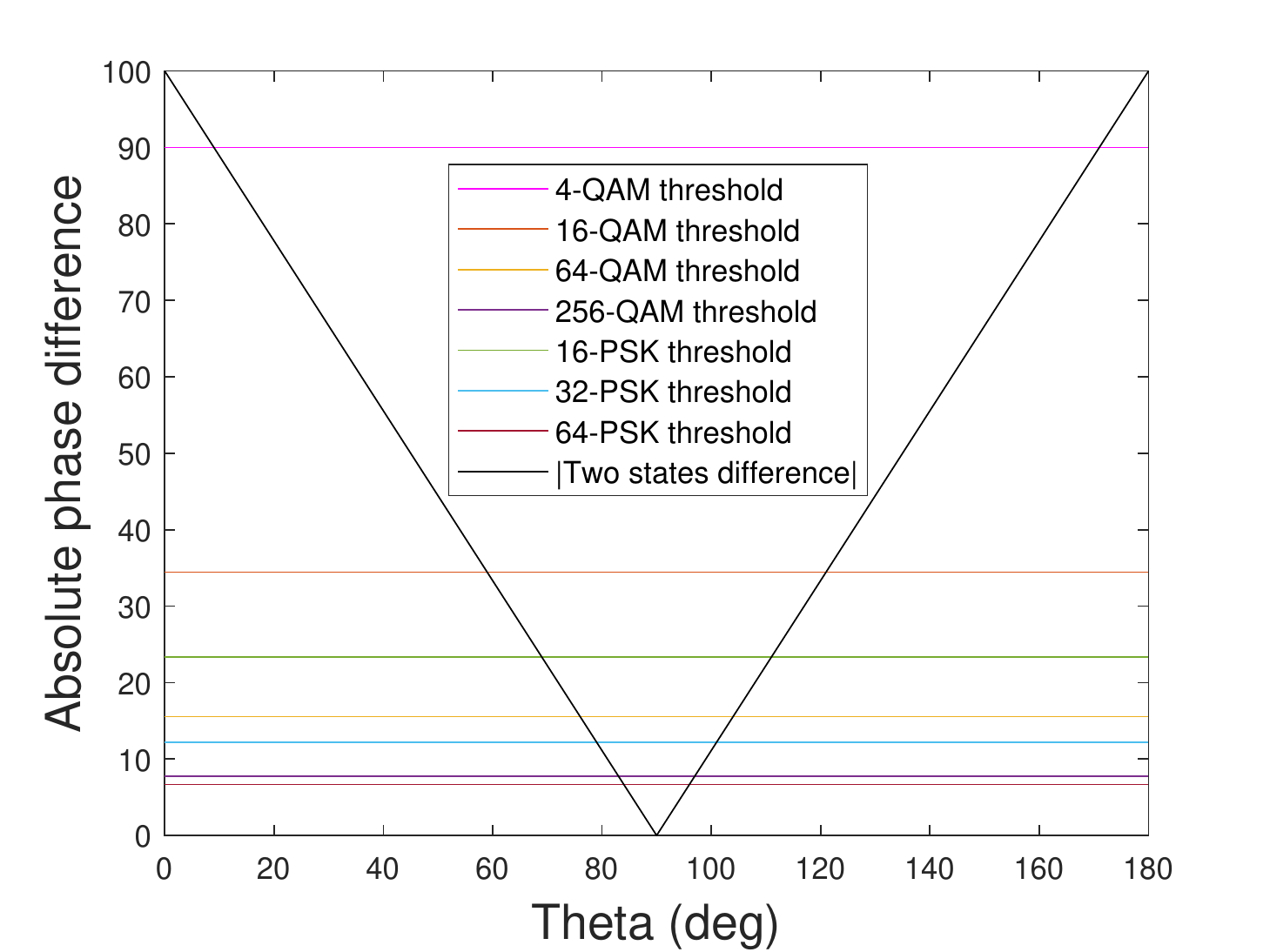}
        \label{Phase_Catalog}
    }
    \caption{Bit error ratio using phase dynamics with no amplitude dynamics and a noise-free channel. The receiver is assumed to know the bit rate and modulation format. (a) BER versus angle for various QAM and PSK orders showing that larger orders are more affected and yield a narrower secure region. (b) The difference between the two phase patterns with thresholds lines below which the corresponding modulation schemes exhibit low BER.}
   \label{Phase_Dynamics_BER}
\end{figure}

\begin{figure}
    \centering
     \subfigure[]
    {
        \includegraphics[width=0.48\textwidth]{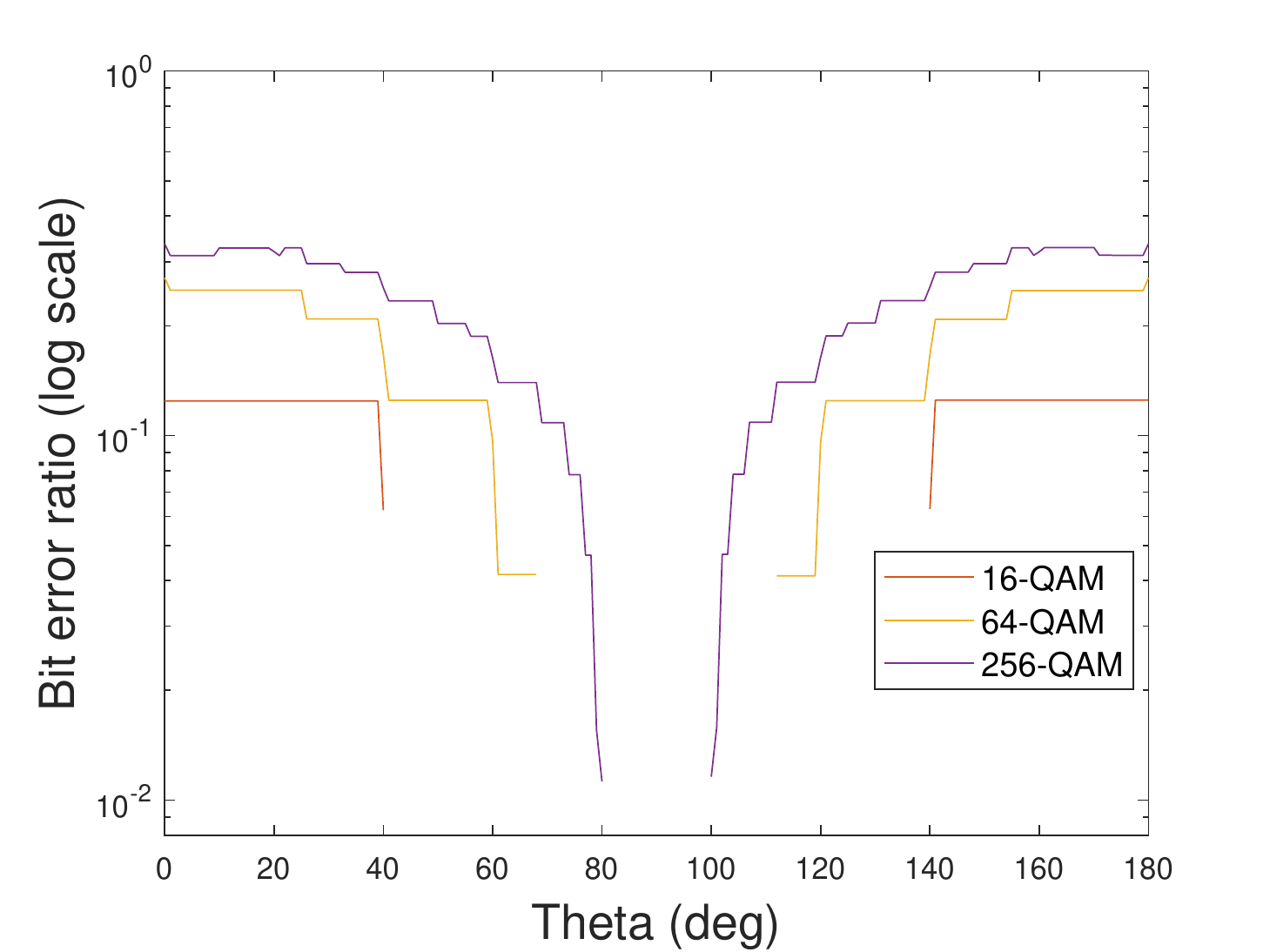}
        \label{AmpBER}
    }
    \\
    \subfigure[]
    {
        \includegraphics[width=0.48\textwidth]{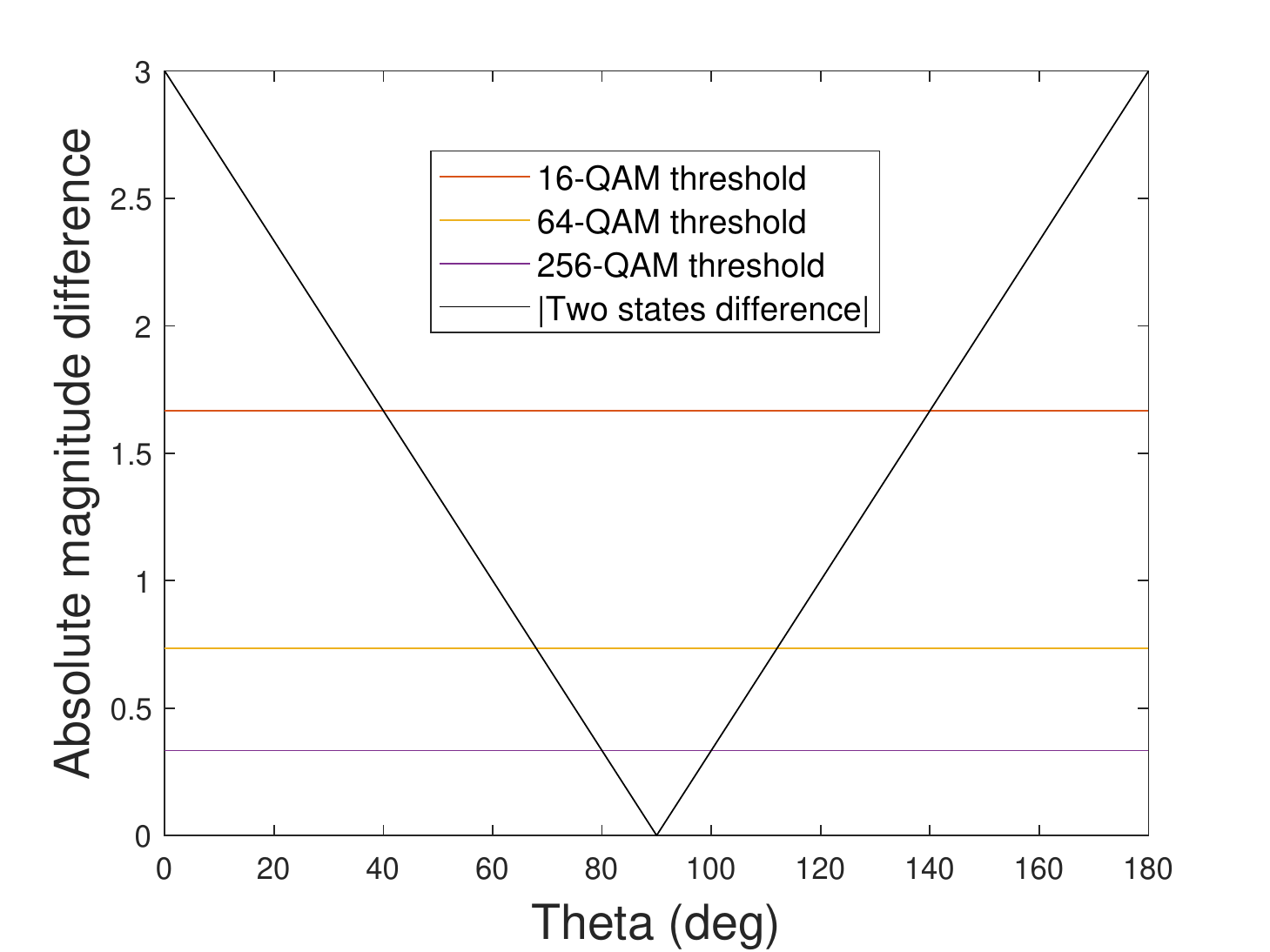}
        \label{Amp_Catalog}
    }
    \caption{Bit error ratio using amplitude dynamics with no phase dynamics and a noise-free channel. The receiver is assumed to know the bit rate and modulation format. (a) BER versus angle for various QAM orders showing that larger orders are more affected and yield a narrower secure region. (b) The difference between the two amplitude patterns with thresholds lines below which the corresponding modulation schemes exhibit low BER.}
   \label{Amplitude_Dynamics_BER}
\end{figure}

In this section, we perform a numerical assessment in MATLAB to assess our theoretical technique for directional modulation introduced in Section~\ref{Theory}. We analyze through simulation the bit-error ratio (BER) obtained as a function of angle for both phase and amplitude dynamics. We simulated PRBS signals with Gray coding and calculated the BER as a function of angle with the effect of the dynamic antenna patterns included. Standard forward error correction (FEC) codes can reliably recover information when BER $<10^{-3}$, thus it is desired that the BER in the secure region be below this threshold, while outside the secure region it should be greater. Fig.~\ref{Phase_Dynamics_BER}(a) and Fig.~\ref{Amplitude_Dynamics_BER}(a) show the resultant BER for noise-free channels for different modulation rates using phase dynamics and amplitude dynamics, respectively. We assume the worst case where the eavesdropper knows the modulation order and the bit rate. It can be seen that a secure region is obtained for each modulation format (centered at $90\degree$) and that higher order modulation formats incur more errors away from the secure region, as expected due to the decreased separation between the symbols. Note that since there is no noise, the increase in BER outside the secure region is due solely to the modulation imparted by the antenna element dynamics. Furthermore, the addition of noise will thus serve to narrow the secure region further. 

Figs. \ref{Phase_Catalog} and \ref{Amp_Catalog} show the difference between the two states in the phase dynamics and amplitude dynamics cases, respectively, in addition to a threshold line for each modulation scheme below which zero BER is obtained. In other words, these threshold lines indicate the minimum amount of difference in phase or amplitude required to cause an error. 
These threshold lines show how the secure region width (the information beamwidth) can be altered by changing the modulation order. 

\section{Switched Dipole Antenna Design}
\label{Design}

\begin{figure}[t!]
\centering
\includegraphics[width=2.75in]{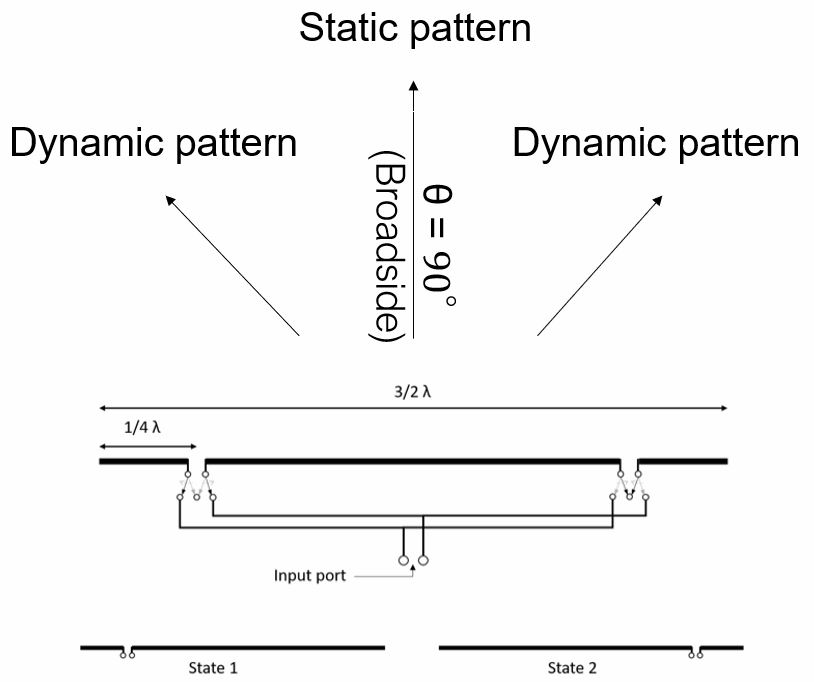}
\caption{The basic concept of dynamic pattern $3\lambda/2$ dipole antenna. The antenna switches between two states that mirror each other; with each state generating an asymmetric radiation pattern. Thus, the pattern in the mainlobe is static while it is dynamic elsewhere.}
\label{Dipole_Fig}
\end{figure}

The above analysis shows the impact on wireless transmission from amplitude or phase dynamics separately; in practice, the amplitude and phase patterns of an antenna are coupled, thus the goal is to obtain a pattern that is static in both phase and amplitude in the broadside direction when switched between two states, and that exhibits dynamics in phase and/or amplitude at angles away from broadside. Our design is based on an asymmetrically-fed 2.26 GHz $3\lambda/2$ dipole antenna whose input signal is switched between two symmetric points on the antenna at distance of $\pm\lambda/2$ with respect to the center (see Fig. \ref{Dipole_Fig}). By switching between two ports and shorting the current path on the unused port, the current distribution is mirrored between the two antenna states; in theory, the antenna pattern will thus remain static at broadside and will change at angles away from broadside. 

\begin{figure}[t!]
\centering
\includegraphics[width=0.48\textwidth]{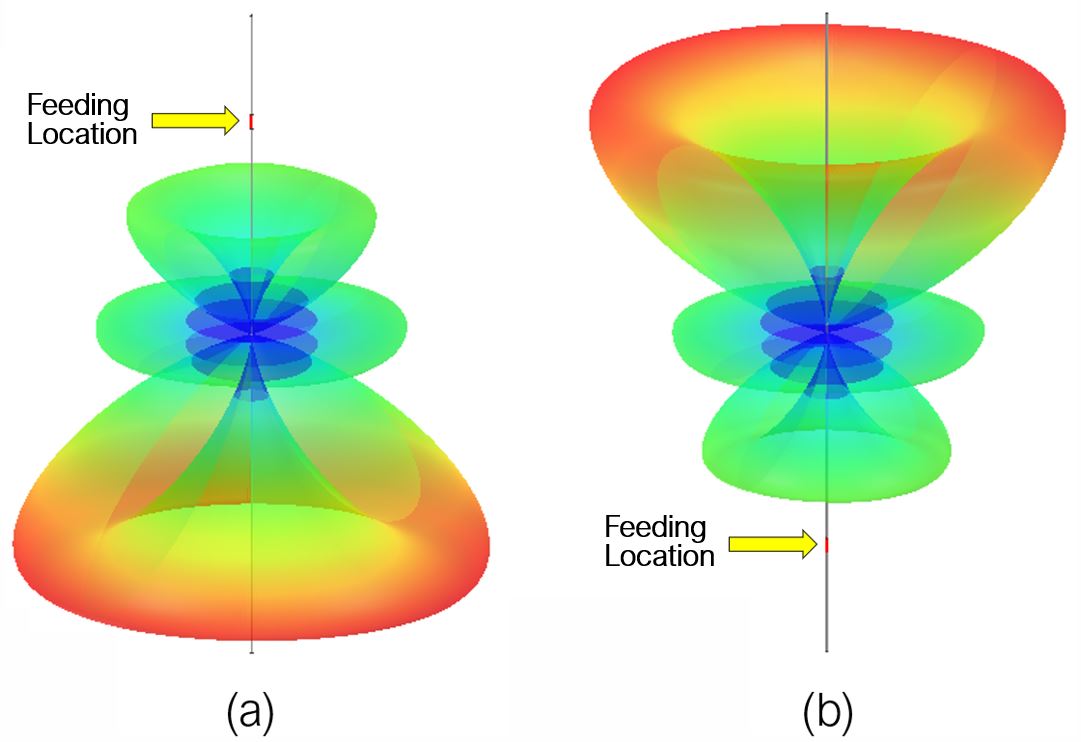}
\caption{(a) Radiation pattern (linear scale) of a $3\lambda/2$ dipole antenna fed at $+\lambda/2$ and (b) fed at $-\lambda/2$ with respect to the center.}
\label{RP_wire}
\end{figure}

Fig. \ref{RP_wire} shows the radiation patterns of the two states; since the feed points of the two states are symmetric, generating mirrored current densities, the resultant radiation patterns are likewise mirrored. Since the antenna is long compared to a wavelength, multiple lobes manifest. The lobe at broadside is constant between the two states, while the two adjacent sidelobes exhibit large differential amplitudes. Such an antenna could feasibly be used for single-element directional modulation, however with reduced gain in the mainbeam.

\begin{figure}[t!]
\centering
\includegraphics[width=0.48\textwidth]{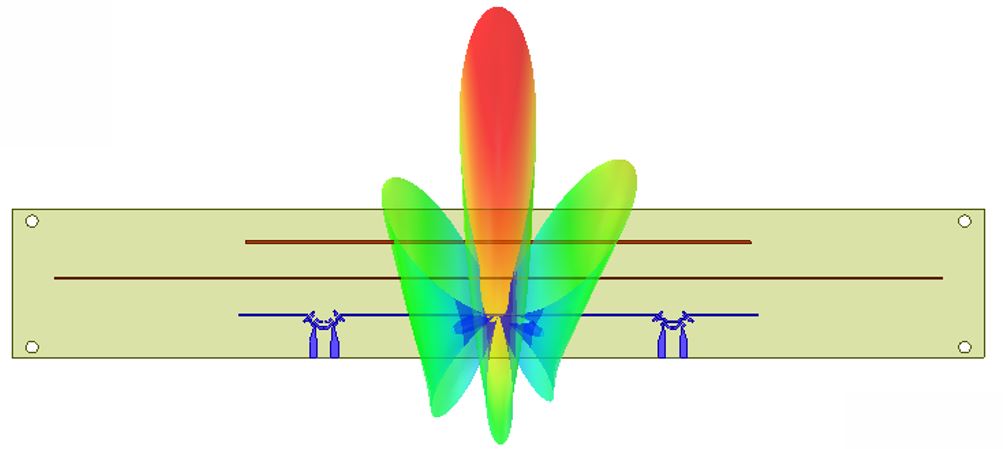}
\caption{Printed dipole design with two parasitic elements and its radiation pattern plotted at 2.26 GHz (linear scale) when fed on the left port. The central lobe is the largest and radiation is mainly in the forward direction, broadside to the dipole, with amplitude variation remaining on the sidelobes.}
\label{static_dipole}
\end{figure}


\begin{figure}[t!]
\centering
\includegraphics[width=0.48\textwidth]{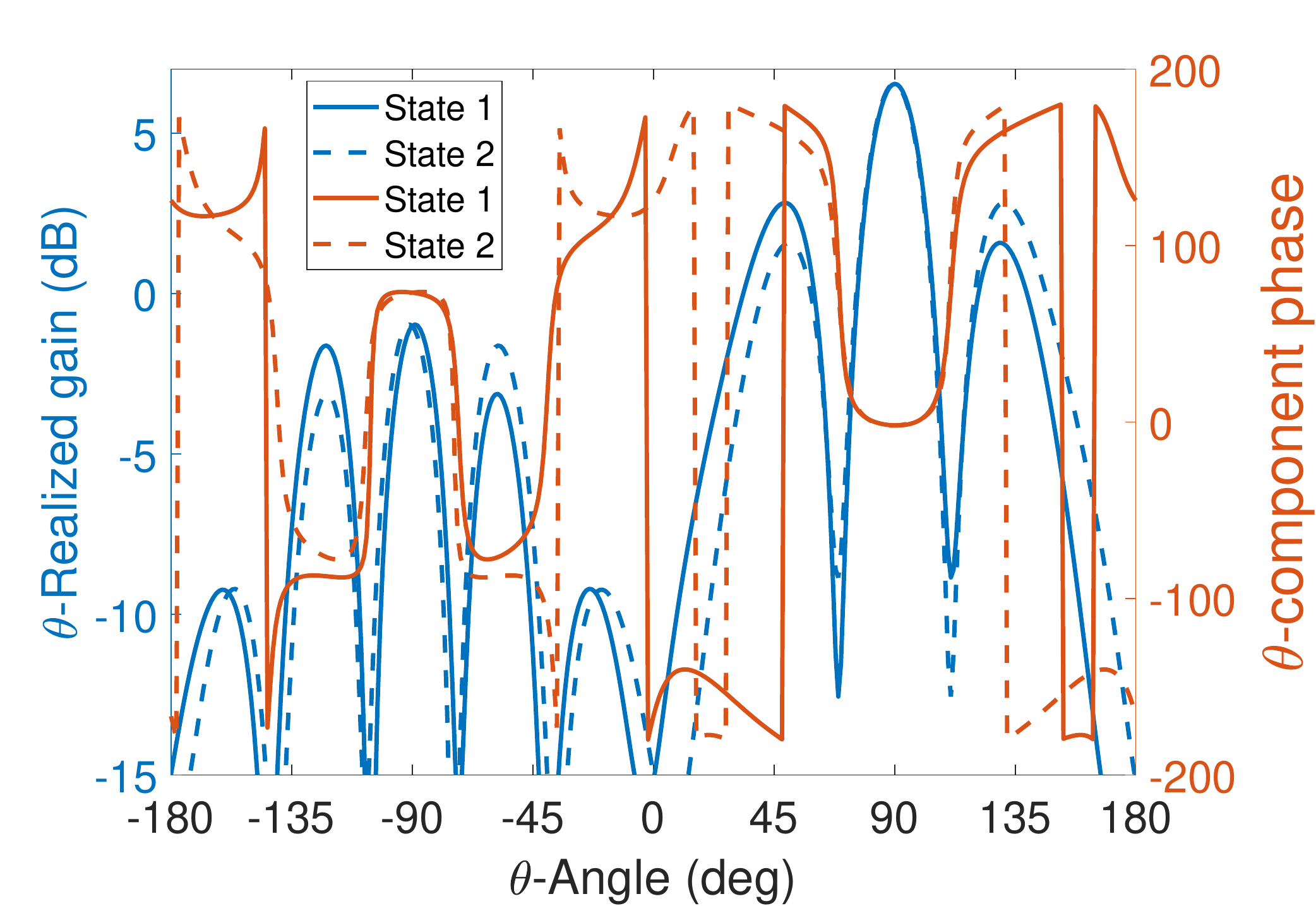}
\caption{Amplitude pattern and phase pattern of the simulated static printed dipole at 2.26 GHz. The central forward region has a quasi-static fields in addition to other three regions in the backside.}
\label{Two_States}
\end{figure}

\begin{figure}[t]
\centering
\includegraphics[width=0.5\textwidth]{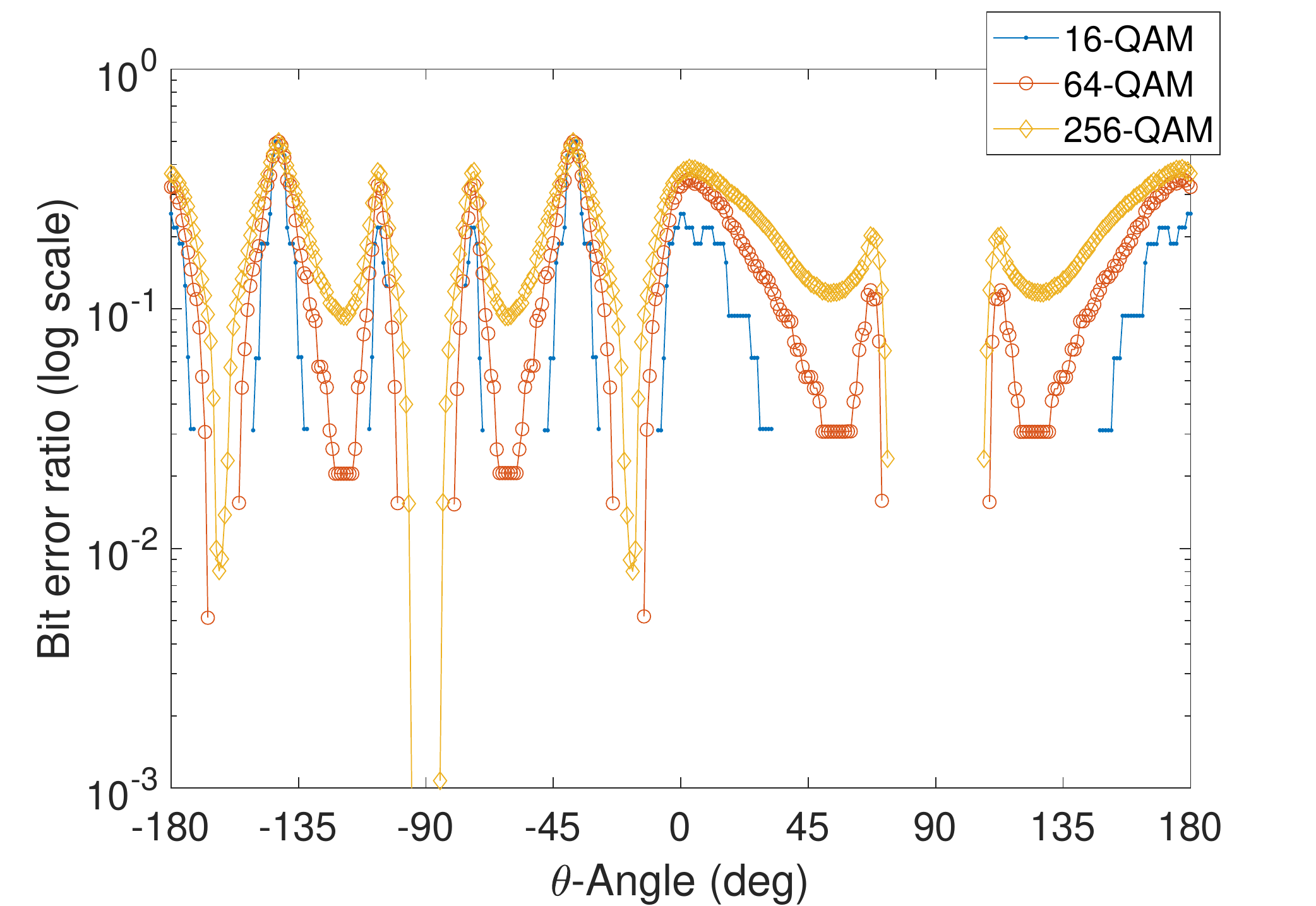}
\caption{BER in a noise-free channel using the amplitude pattern and phase pattern shown in Fig. \ref{Two_States}.}
\label{BER_HFSS}
\end{figure}


To overcome the reduced gain in the broadside beam, two parasitic directive elements were added, and the antenna was designed to be printed on a Rogers RO4350B dielectric substrate with thickness of 1.524mm, as shown in Fig. \ref{static_dipole}.  
The antenna uses a balanced feed with coplanar striplines 
~\cite{jamaluddin1,jamaluddin2}. 
The length of the dipole radiator is 148.08\;mm, the longer parasitic element is 253.33\,mm, and the shorter parasitic element is 144.08\;mm. The width of each line is 0.55\;mm and the distance between the center of any two consecutive elements is 10.37\;mm. The feeding lines have a length of 4.5\;mm (before tapering), a width of 2\;mm, and their centers are separated by 6.19mm. The length and width of the substrate are 277.08\;mm and 42.83\;mm. 

The radiation pattern overlaid on the antenna is the result of feeding the left port. The realized gain and phase patterns for both states is plotted in Fig. \ref{Two_States}. The addition of the directors has a significant impact on the center lobe and sidelobes, as most of the signal energy is now constrained to the center lobe with a large front-to-back ratio (FBR). However, there remains an amplitude imbalance between the two sidelobes; this is necessary to ensure that the radiation pattern in the sidelobes is dynamic when switching between the two states, as the radiation pattern will be mirrored for the other state.

We simulated the BER in a noise-free channel, assuming the eavesdropper knows the bit rate and modulation order, using the amplitude and phase patterns from the simulated antenna. 
The corresponding BER curves are shown in Fig. \ref{BER_HFSS}. 
Note that there are quasi-static fields in the central forward lobe, and also in three directions in the backward region. Note that secure region in the forward direction is not greatly reduced when moving from 64-QAM to 256-QAM because there is no appreciable amplitude or phase dynamics in this region as shown in Fig. \ref{Two_States}. Note also that we have three other secure regions in the backward central direction; however, the realized gain there is small, thus the addition of noise in the simulation will remove these areas of low BER.

\begin{figure}[t!]
\centering
\includegraphics[width=3.5in, angle =0]{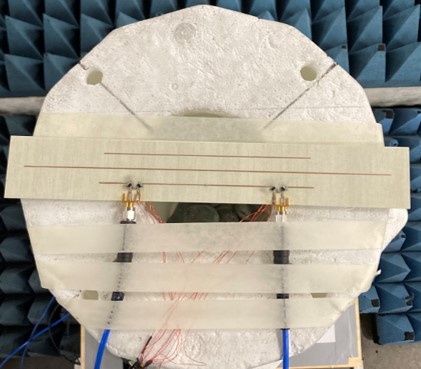}
\caption{Fabricated switched dipole antenna using four switches. The wires are used to control the switches by a microcontroller (MCU). Either one of the two feeding lines carry the signal at a time with an external switch changing the signal feed between them. This external switch is also controlled by the MCU to form either State 1 or State 2.}
\label{Antenna_pic}
\end{figure} 

\begin{figure}[t!]
\centering
\includegraphics[width=0.48\textwidth]{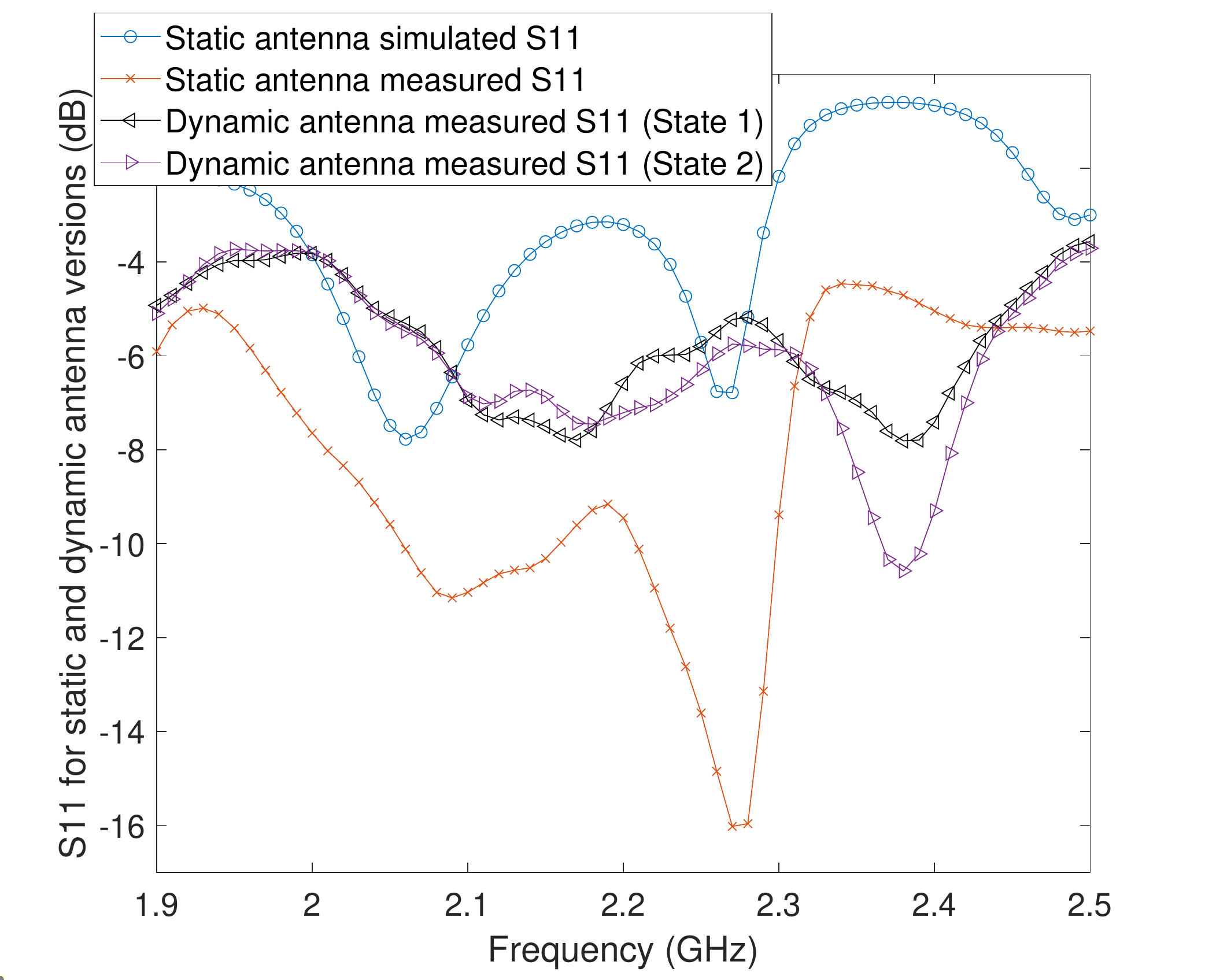}
\caption{S11 of simulated static antenna, printed static antenna, and the dynamic printed dipole.}
\label{S11_figure}
\end{figure}





\section{Dynamic Antenna Fabrication and Measurement}
\label{RP_Measurements}


A static version of the printed dipole in Fig. \ref{static_dipole} was initially fabricated to evaluate the radiation pattern; the State 1 (left) feedport was connected to an SMA connector while the second feedport was shorted. This antenna resonated at 2.27 GHz, close to the simulated resonance frequency. The radiation pattern was measured at 2.275GHz and exhibited a realized gain of approximately 1.5 dBi and a FBR of approximately 6 dB.
The switched dynamic printed dipole was fabricated using 4 switches (HMC545AE) and 7 capacitors (0603ZA101JAT2A), and is shown in Fig. \ref{Antenna_pic}. There is one signal feeding port that feeds the antenna system through an external switch (HMC595A) which connects it to either port 1 feed line (to form State 1) or port 2 feed line (to form State 2). This external switch in addition to the four switches on the antenna are controlled by a microcontroller (MCU). 
The simulated and measured S11 of the single-state antenna and that of the two states of the dynamic antenna are shown in Fig. \ref{S11_figure}. 
The simulations were conducted in HFSS and did not include models of the switches on the antenna, instead replacing them with short circuits for the appropriate switch state, thereby adding some mismatch between the simulated and measured designs.
Inclusion of the switches resulted in an increase in the resonance frequency from 2.26 GHz to 2.385 GHz, and furthermore imperfect matching between the switches and the antenna lines resulted in a worse match. 
Nonetheless, the two states are consistent, resonate at the same frequency, and approach --10 dB S11. Future work to improve the match between the switches and the antenna lines would serve to improve ths S11.
However, the calculated gain for this dynamic printed antenna is around --4 dBi, which is due to the mismatch of the four switches existing in the current path. The switches are designed to be connected to 50 ohm feed lines above a ground; a future design will seek to optimize the impedance matching to the switches, which should lead to increased gain closer to that of the static design. 
Nevertheless, our design is sufficient to demonstrate the technique and show that it can work in short distance applications such as WiFi.

\section{Experimental Demonstration in a Wireless Communication System}

\begin{figure}[t!]
\centering
\includegraphics[width=3.5in]{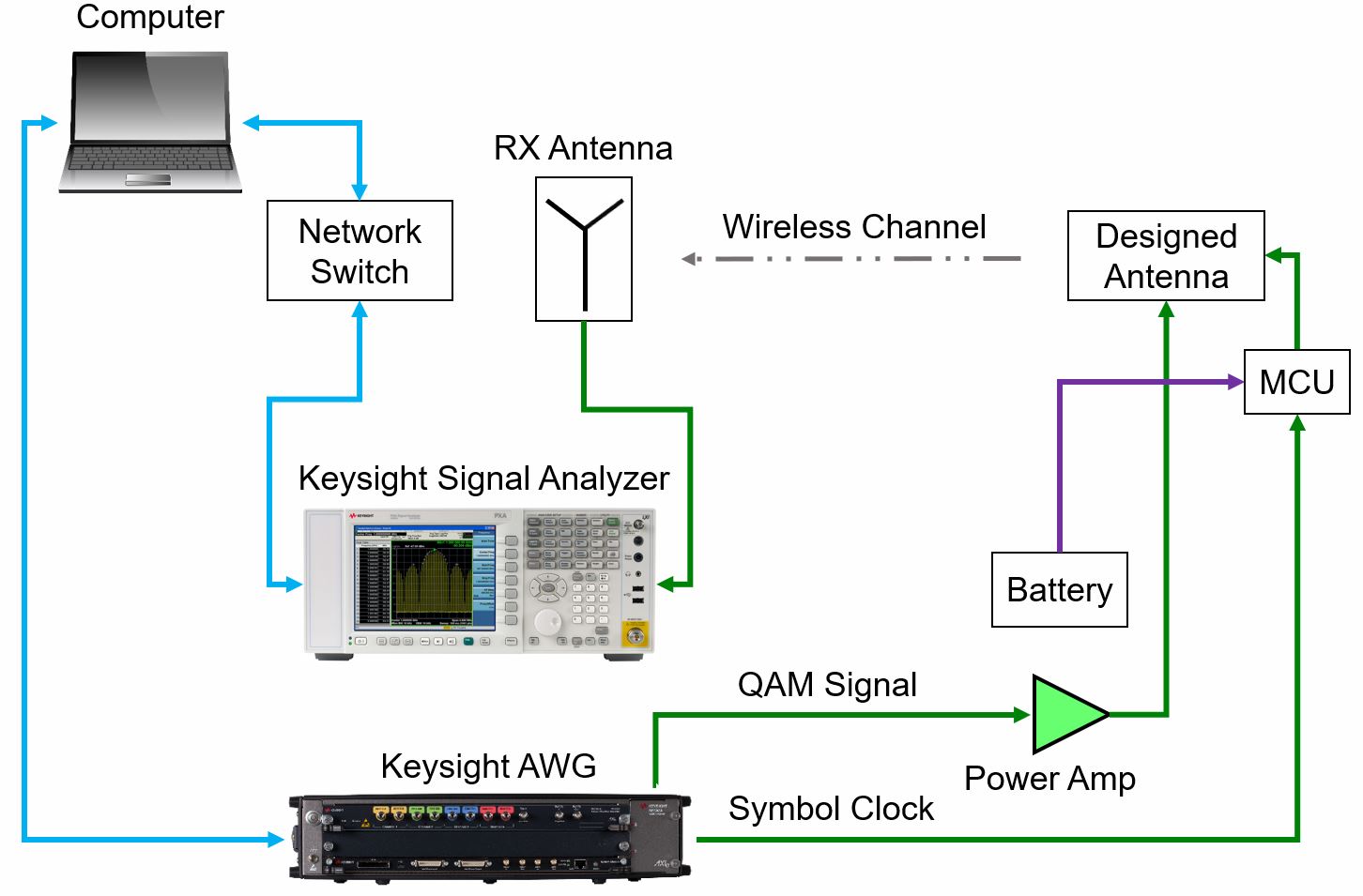}
\caption{Experimental communications system setup. The AWG transmitted data through an amplifier to the antenna which is switched synchronously with the symbol clock; the receiver (RX) send the signal to the signal analyzer.}
\label{system}
\end{figure}


\begin{figure*}
    \centering
    \subfigure[]
    {
        \includegraphics[width=0.7\textwidth]{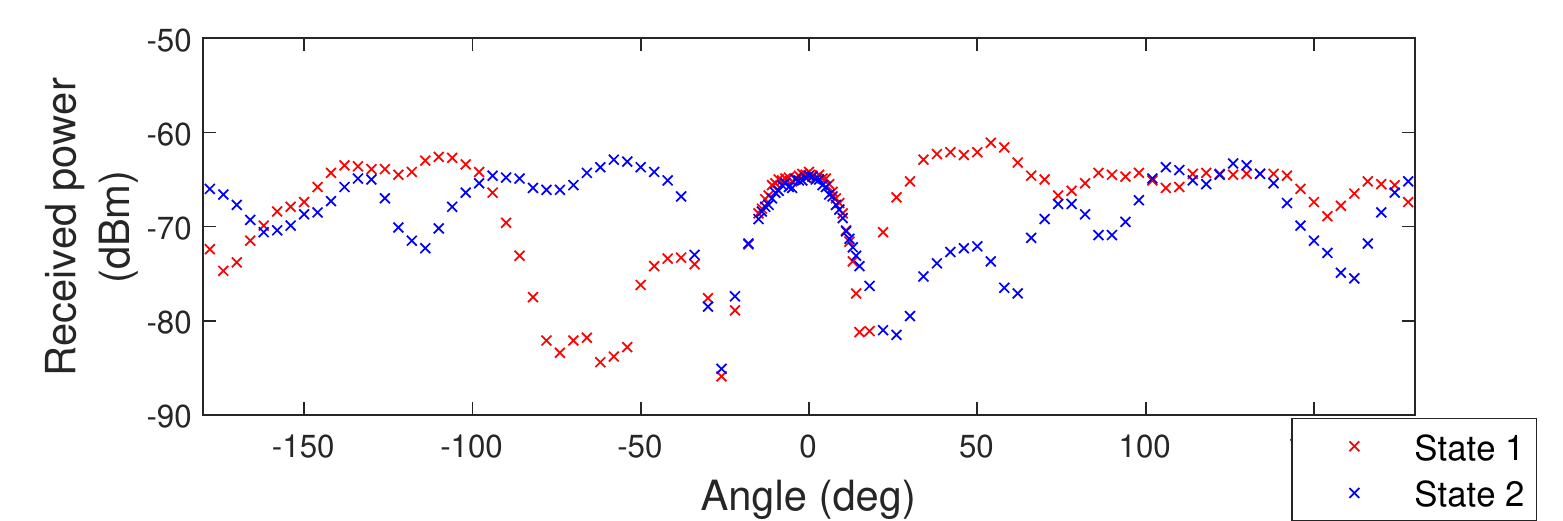}
       \label{Fig:Qa}
    }
    \\
    \subfigure[]
    {
        \includegraphics[width=0.7\textwidth]{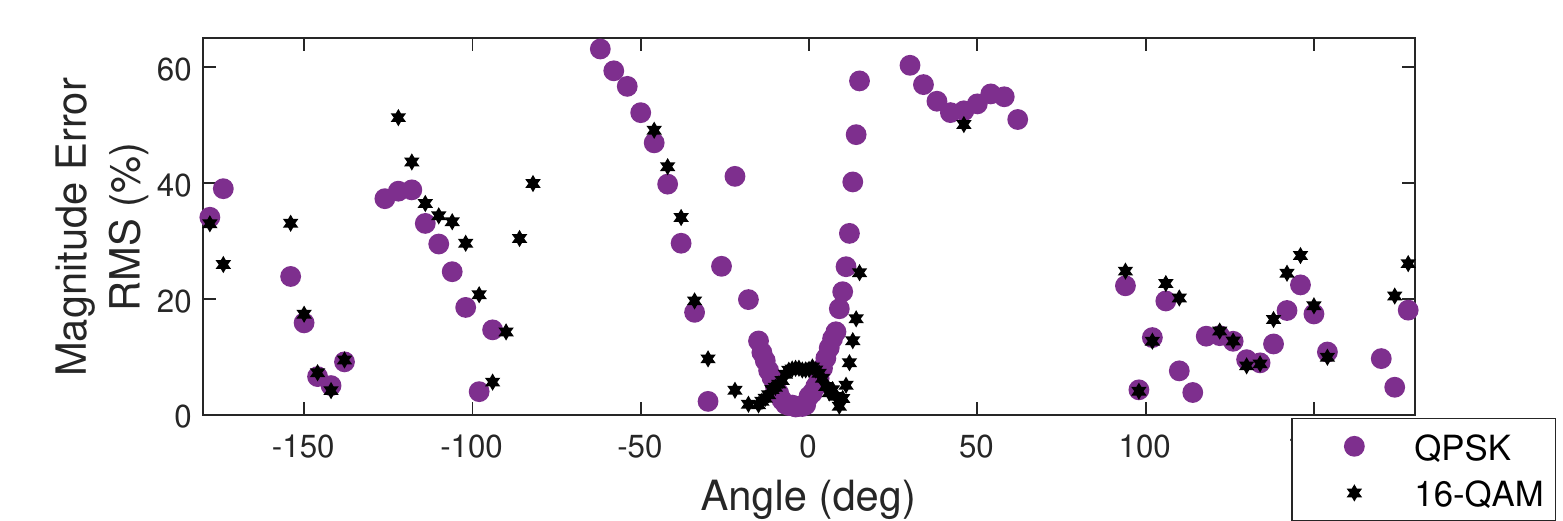}
       \label{Fig:Qb}
    }
    \\
     \subfigure[]
    {
        \includegraphics[width=0.7\textwidth]{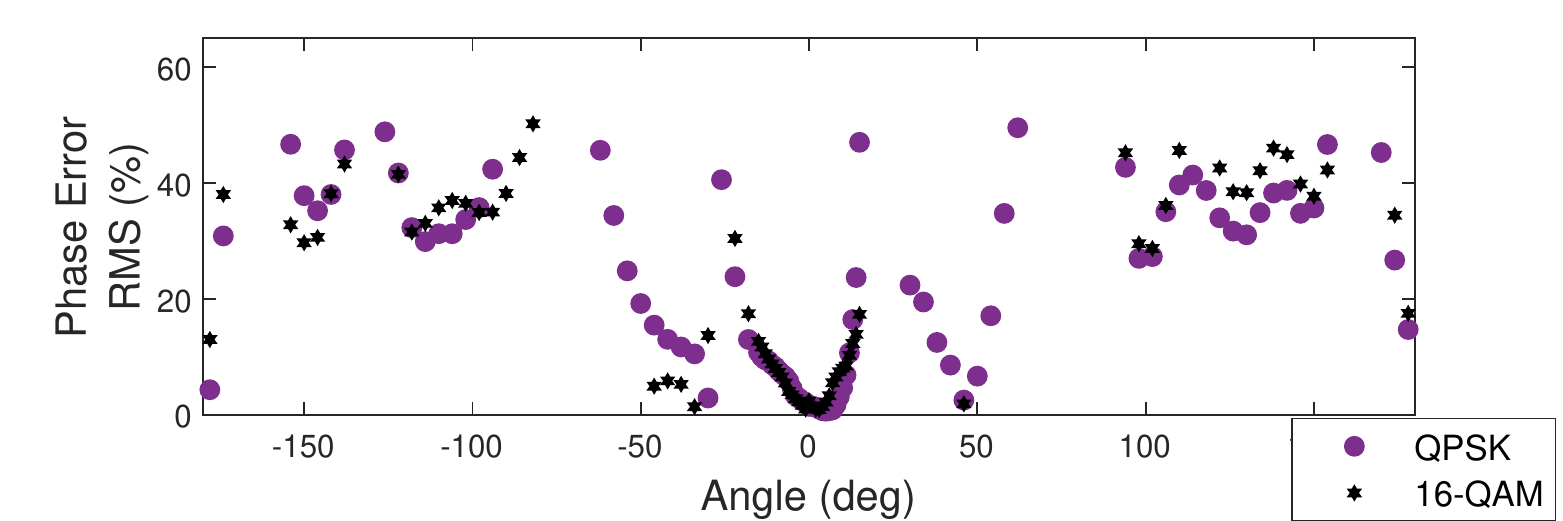}
        \label{Fig:Qc}
    }
    \\
    \subfigure[]
    {
        \includegraphics[width=0.7\textwidth]{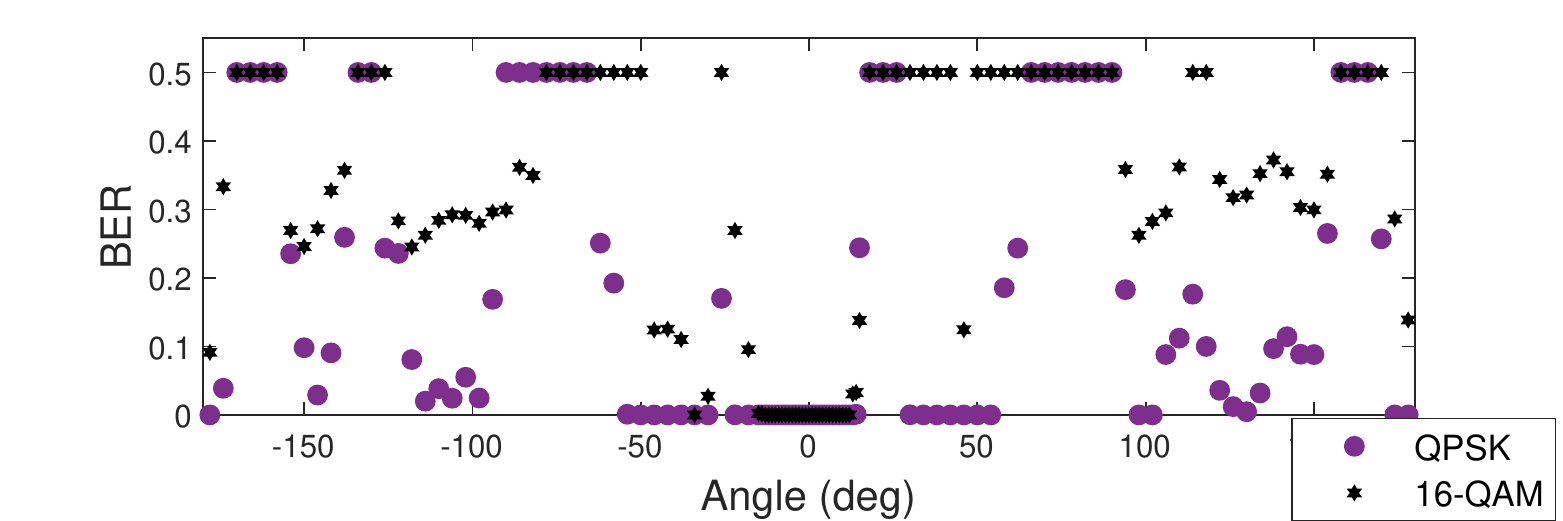}
        \label{Fig:Qd}
    }
    \caption{Measured radiation patterns and error parameters of the dynamic switched antenna when transmitting QPSK or 16-QAM at 2.385 GHz with 1 MHz symbol rate. (a) Radiation patterns of the two static states. (b) Magnitude error during switching. (c) Phase error during switching. (d) BER during switching. The antenna was rotated in a one-degree step in the region from --15$\degree$ to +15$\degree$ to increase sampling in the secure region, while four-degree jumps were taken outside this region (--18$\degree$ and above +18$\degree$). The magnitude error and phase error provide a measure for the radiation pattern distortion due to switching; they are related to the BER: large errors are obtained when either is large. The measurement setup and environment can be considered almost noise free since no errors appeared for the static measurements, even at the nulls and low power regions of the two static states radiation patterns. Some angles are left empty since the signal was so distorted that no data could be demodulated; at these angles, the BER was set to 0.5. Thus, all errors detected were due solely to dynamic switching of the antenna.}    
\label{fig:Measured_errors_parameters}
\end{figure*}


\subsection{System Design}
The block diagram of the experimental system is shown in Fig. \ref{system}. A communications signal is generated directly at a 2.385 GHz carrier frequency using a Keysight M8190A Arbitrary Waveform Generator (AWG) then amplified and passed to the antenna under test. 
The antenna transmits the signal while it is dynamically switched between the two states by the MCU, synchronously with the symbol clock. 
The radiated signal is received by a HG2458-08LP-NF log periodic antenna and sampled in a Keysight N9030A PXA Signal Analyzer. The PC used Keysignt IQtools and PathWave Vector Signal Analysis software to generate and demodulate the communications signals in real time. 

\begin{figure*}[t!]
    \centering
    \subfigure[]
    {
        \includegraphics[width=0.25\textwidth]{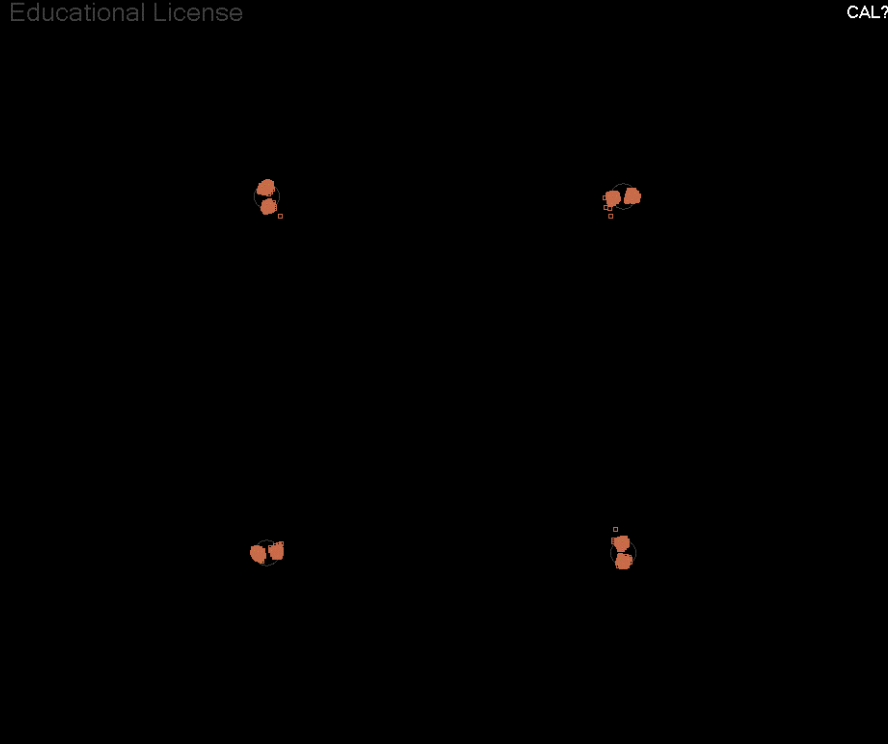}
        \label{QPSK_0_constellation}
    }
    \subfigure[]
    {
        \includegraphics[width=0.25\textwidth]{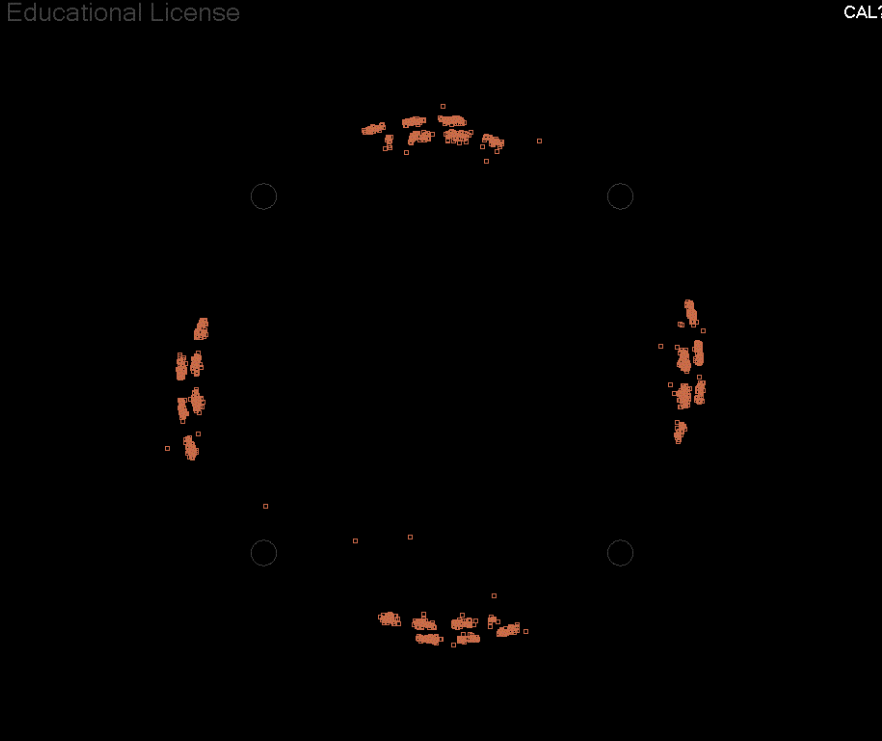}
       \label{QPSK_114_constellation}
    }
    \subfigure[]{
        \includegraphics[width=0.25\textwidth]{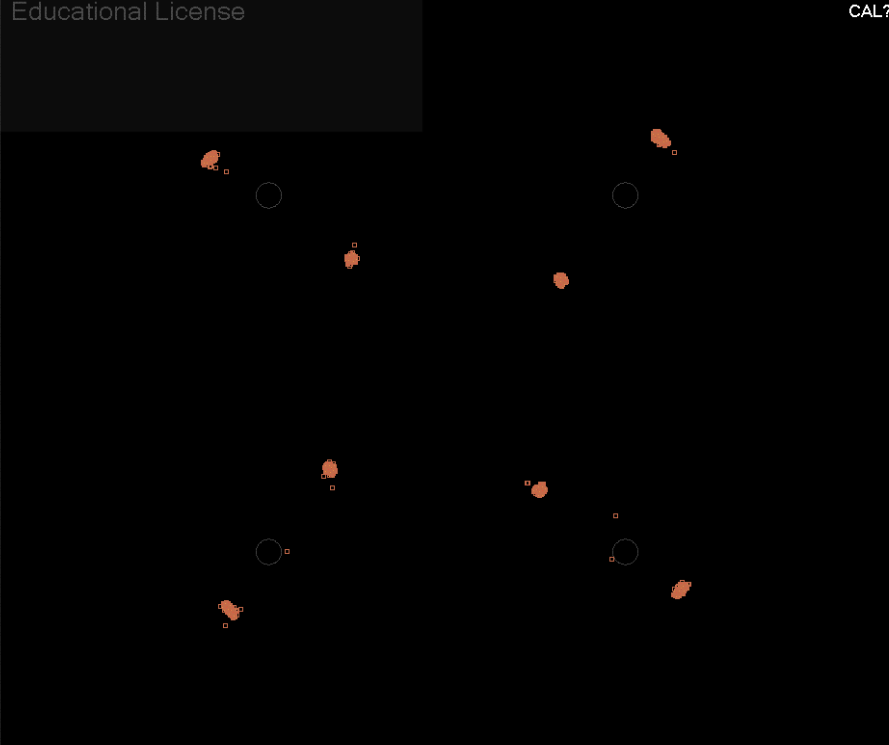}
        \label{QPSK_neg_178_constellation}
    }
    \\
    \subfigure[]
    {
        \includegraphics[width=0.25\textwidth]{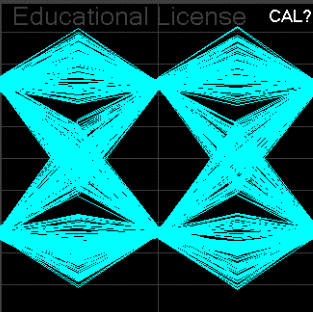}
        \label{QPSK_0_eye}
    }
     \subfigure[]
    {
        \includegraphics[width=0.25\textwidth]{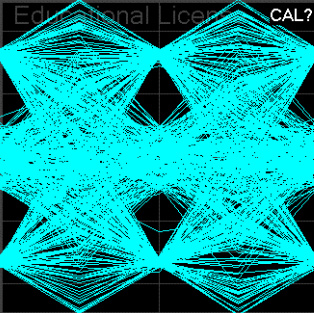}
       \label{QPSK_114_eye}
    }
    \subfigure[]
    {
        \includegraphics[width=0.25\textwidth]{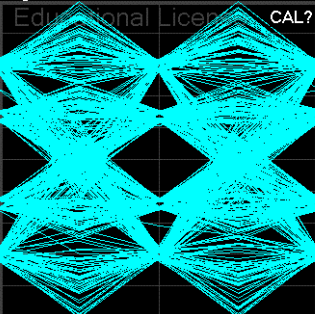}
    \label{QPSK_neg_178_eye}
    }
    \caption{Measured constellations and eye diagrams for QPSK signals at 0$\degree$ [(a) and (d)], 114$\degree$ [(b) and (e)], and --178$\degree$ [(c) and (f)]. (a) and (d) correspond to BER = 0; (b) and (e) correspond to BER = 0.176; (c) and (f) correspond to BER = 0.}
    \label{QPSK_samples}
\end{figure*}

\begin{figure*}[t!]
    \centering
    \subfigure[]
    {
        \includegraphics[width=0.25\textwidth]{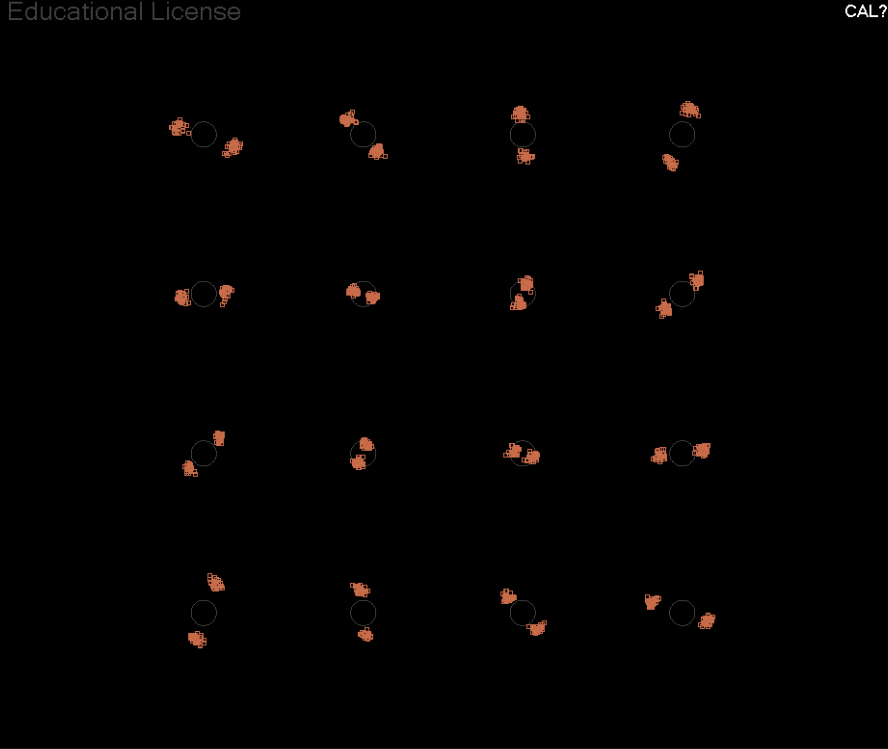}
       \label{QAM16_0_constellation}
    }
        \subfigure[]
    {
        \includegraphics[width=0.25\textwidth]{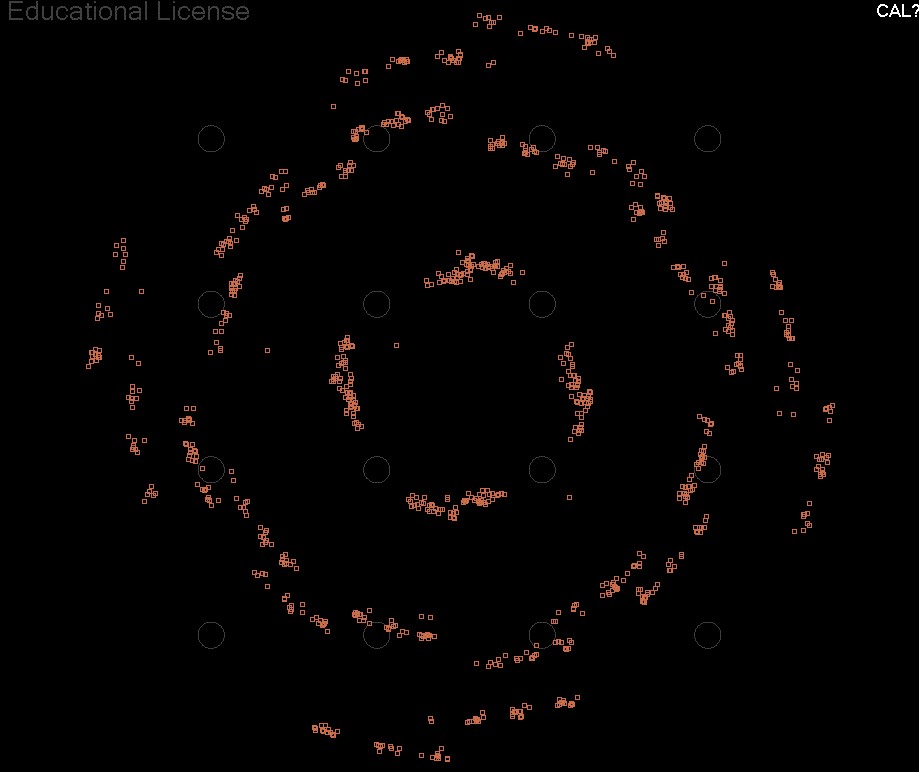}
       \label{QAM16_neg_94_constellation}
    }
     \subfigure[]
    {
        \includegraphics[width=0.25\textwidth]{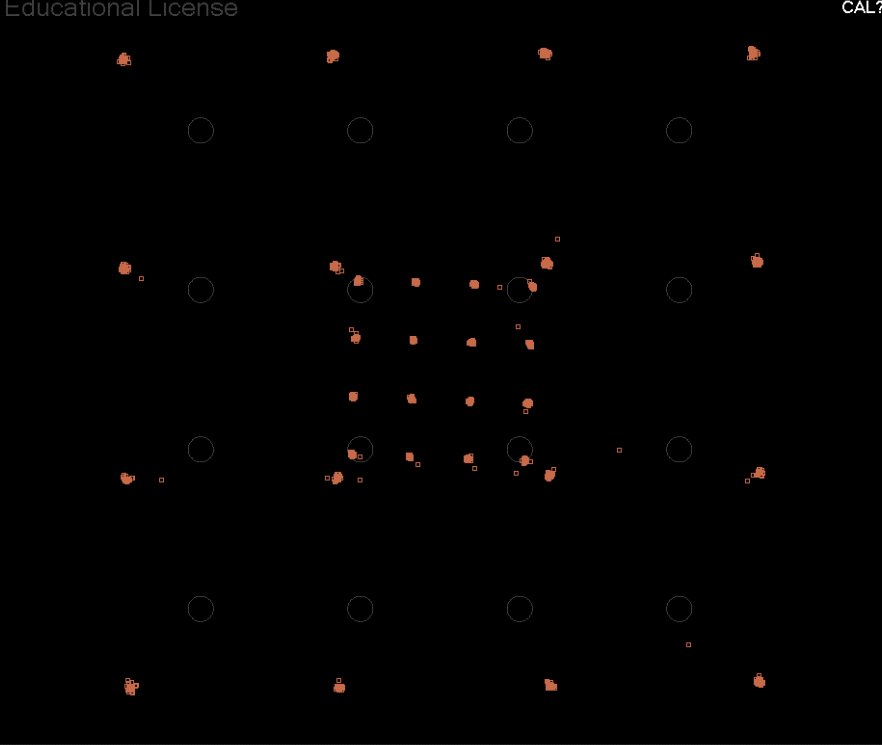}
       \label{QAM16_46_constellation}
    }
    \\
    
    \subfigure[]
    {
        \includegraphics[width=0.25\textwidth]{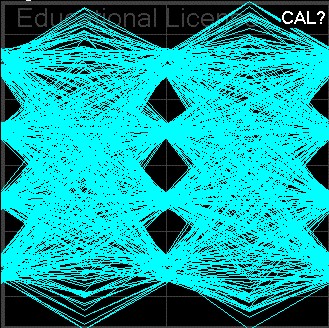}
       \label{QAM16_0_eye}
    }
    \subfigure[]
    {
        \includegraphics[width=0.25\textwidth]{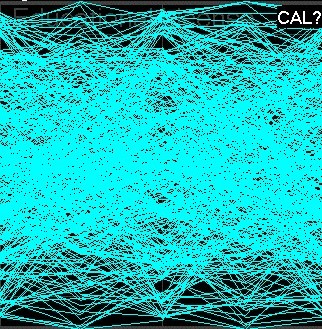}
       \label{QAM16_neg_94_eye}
    }
     \subfigure[]
    {
        \includegraphics[width=0.25\textwidth]{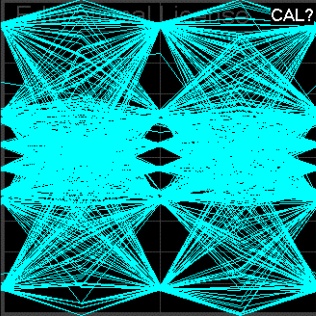}
       \label{QAM16_46_eye}
    }
    \caption{Measured constellations and eye diagrams for 16-QAM signals at angles 0$\degree$ [(a) and (d)], --94$\degree$ [(b) and (e)], and 46$\degree$ [(c) and (e)]. (a) and (d) correspond to BER = 0; (b) and (e) correspond to BER = 0.296; (c) and (f) correspond to BER = 0.124.}
    \label{QAM16_samples}
\end{figure*}

The dynamic antenna was calibrated for quasi-static amplitude and phase among the two states at broadside angle before taking each dynamic measurement. Thus, a phase shifter was connected to one of the cables going to port 1 to align the phases between the two states, and a 6-dB attenuator was connected to port 2 to match the amplitudes of the two states. These imbalances were due largely to differences in the cables and losses in the switches and phase shifter; in an integrated design the imbalances would be less and also easier to correct with small changes in microstrip lengths, etc.

The antenna was measured in both a static and dynamic cases in a semi-enclosed anechoic environment at a center frequency of 2.385 GHz. The antenna was rotated in one-degree increments in the region from -15$\degree$ to +15$\degree$, which is in the main beam, and in four-degree increments outside this region. The measured power from the two states of the antenna when driven statically (no switching) is shown in Fig.~\ref{fig:Measured_errors_parameters}(a), which clearly shows the consistency of the main beam between the two states, and the difference in the sidelobe power between the two states. For measuring the BER as a function of angle, $2^{11}-1$ PRBS signals of length 30,000 symbols were transmitted from the dynamic antenna with QPSK and 16-QAM modulation formats at a symbol rate of 1 MSps. 48Kbits were used to compute the error metrics at each angle. Demodulation was performed assuming knowledge of the transmitter parameters such as modulation format and symbol rate, representing a worst-case eavesdropper.

\subsection{Measured Results vs. Angle}
\label{BER_Measurements}
The measured magnitude error, phase error, and BER versus angle are shown in Figs. \ref{Fig:Qb}, \ref{Fig:Qc} and \ref{Fig:Qd}, respectively. 
The magnitude error is the RMS average of the difference in amplitude between the I/Q measured signal and the I/Q reference signal.
The phase error is the phase difference between the I/Q reference signal and the I/Q measured signal, averaged over all symbol points.
In these three plots, and for some of the angles where distorted signals appear, then there was nothing detected in the reception side which most likely indicates that the signal was distorted to the level that reception was not possible. These regions were left empty in Figs. \ref{Fig:Qb} and \ref{Fig:Qc} while they were replaced with the value 0.5 in the measured BER plot (Fig. \ref{Fig:Qd}) as 0.5 is the worst case BER value. 
The magnitude error (Fig. \ref{Fig:Qb}) is consistent with the differences in the received power between the two states as indicated in Fig. \ref{Fig:Qa}: very small error is seen near broadside, indicating that the dynamic switching has a negligible effect on the transmitted data, whereas at other angles the error is high due to the additional modulation imparted on the signal from dynamic switching.
Similarly, the phase error (Fig. \ref{Fig:Qc}) is small in the mainbeam and appreciable elsewhere. 
Note that although the phase error is low in the two sidelobes adjacent to the mainbeam, the magnitude error is high; since either amplitude or phase dynamics will impart directional modulation, the signal will be modulated in amplitude in the adjacent sidelobes.
This is clearly shown in the measured BER (Fig. \ref{Fig:Qd}). The QPSK case exhibits zero BER (or very small value $< 10^{-3}$ at various angles that are not limited to the broadside direction. However, QAM-16 case only reliably has BER $< 10^{-3}$ in the region from -12$\degree$ to 13$\degree$. This demonstrates the theory described earlier that large modulation orders are more affected by magnitude and phase dynamics.


Measurements of static antennas (in only States 1 or 2 without switching) were conducted by measuring the BER at the nulls and low power regions shown in Fig. \ref{Fig:Qa}, when transmitting 16-QAM signals (16-QAM is more susceptible to errors compared to QPSK signals). All of these regions yielded zero BER. A static antenna exhibits errors mainly in nulls and other low gain regions where the SNR is poor; thus, since the BER was zero in the static case even at the nulls, all errors generated in the dynamic case were the result of dynamic switching, and not low SNR. The measurements thus closely resembled the noise-free channel case assumed in the simulations above.
\subsection{Received Data}
Examples of constellation diagrams and eye diagrams captured at the receiving antenna 
are discussed here for the QPSK and 16-QAM cases to elaborate more on the relation between the three error metrics we just discussed, i.e., magnitude error, phase error, and BER. The samples for QPSK case are shown in Fig. \ref{QPSK_samples} at angles 0$\degree$, 114$\degree$, and --178$\degree$ with the corresponding measured BER. 
At broadside (0$\degree$) the magnitude error and phase error are small, and constellation and eye diagram look standard. At 114$\degree$, there is a small magnitude error and large phase error that resulted in a constellation with four major points, however the symbols here are misplaced as in Fig. \ref{Example}, leading to a distorted eye diagram and high BER (0.176). Another interesting case is shown at angle --178$\degree$ at which there is a large amplitude error and small phase error; however, PSK modulation orders are not affected by amplitude dynamics and thus the measured BER at this angle is zero.

Constellation diagrams and eye diagrams of similar scenarios for the QAM-16 case are provided in Fig. \ref{QAM16_samples}. The angles studied are 0$\degree$, --94$\degree$, and 46$\degree$. For the 0$\degree$ angle then we have a standard QAM-16 signal. 
The --94$\degree$ case has small magnitude error and large phase error leading to distorted diagrams and a BER of 0.296. 
In this case, magnitude and phase error relation is analogous to that of the QPSK case at 114$\degree$, shown in Fig. \ref{QPSK_114_constellation}, since both QPSK and 16-QAM are affected mainly by phase dynamics at this angle. 
Another interesting case occurs at angle 46$\degree$ which has a large magnitude error and small phase error. As expected, this yielded large BER (0.124) since general large QAM modulation schemes are affected by amplitude dynamics.


\begin{figure}
    \centering
     \subfigure[]
    {
        \includegraphics[width=0.22\textwidth]{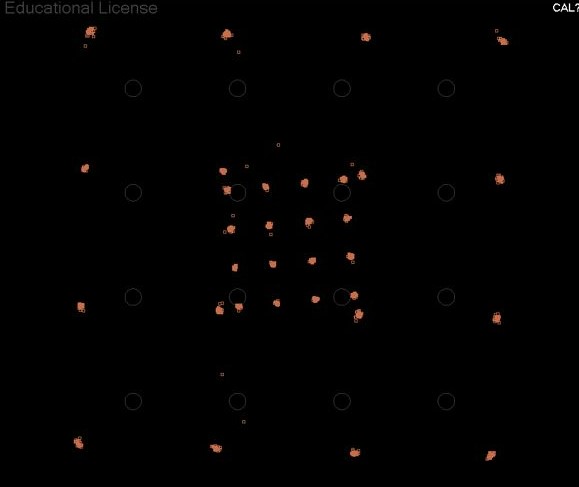}
       \label{Fig:Sa}
    }
         \subfigure[]
    {
        \includegraphics[width=0.22\textwidth]{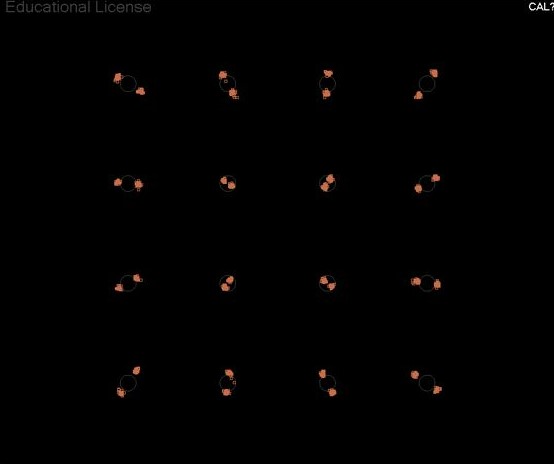}
       \label{Fig:Sc}
    }
    \\
    \subfigure[]
    {
        \includegraphics[width=0.22\textwidth]{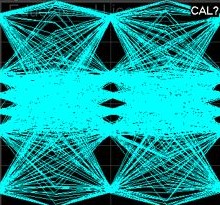}
       \label{Fig:Sb}
    }
    \subfigure[]
    {
        \includegraphics[width=0.22\textwidth]{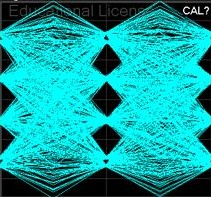}
       \label{Fig:Sd}
    }
    \caption{Measured constellations and eye diagrams for 16-QAM signals where the secure region was steered to 46$\degree$ at angles 0$\degree$ [(a) and (c)] and 46$\degree$ [(b) and (d)], showing a clear constellation at the beamsteering angle and distorted signals at broadside.}
   \label{fig:steering}
\end{figure}

\subsection{Steering the Secure Region}
\label{Steering_Section}

As discussed in Section \ref{Theory}, the secure region can be steered by correcting for the phase and amplitude imbalance between the two states at a desired angle. This can be obtained in practice via a look-up table of the differential complex radiation patterns between the two states. 
By checking the radiation pattern of the two states shown in Fig. \ref{Fig:Qa}, at an angle of 122$\degree$ for example, only phase calibration is needed since there is a small amplitude difference. However, a more general case can be considered by steering the secure region to angle 46$\degree$ which has an amplitude difference of almost 10 dB. By correcting the amplitude and phase imbalance, the secure region was steered to 46$\degree$ as shown in Fig. \ref{fig:steering}. In Fig. \ref{fig:steering}, we see that a distorted constellation diagram and eye diagram appear at 0$\degree$ while standard constellation and eye diagrams are obtained at 46$\degree$, demonstrating the ability to steer the secure region.

 \section{Conclusion}
 \label{Conclusion}
 
A new approach to directional modulation using a dynamic antenna was introduced and demonstrated through a $3/2\lambda$ printed dynamic dipole antenna operating at 2.385 GHz. 
The introduction of a dynamic antenna supports directional modulation in a single antenna element, where prior works have used arrays or multiple signal feeds. The dynamic antenna approach is thus applicable to compact wireless system applications where arrays or the generation of multiple RF input signals is not feasible.
Furthermore, our approach can be implemented as a ``black box'' such that its operation can be transparent to the rest of the wireless system, providing a new form of security at the physical layer that is applicable to both sensing and communications.
In addition, the underlying theoretical concept of using dynamic amplitude and phase patterns for directional modulation is applicable to antennas beyond the single-element design shown in this work.
Our experimental results demonstrated the ability to maintain low BER in an intended direction (both broadside and steered) while mitigating the ability to demodulate the information at other angles.
The gain of the dipole antenna can be further improved by better matching of the switches to the antenna elements, and a more detailed implementation of directional elements can potentially improve overall gain and FBR. 

\ifCLASSOPTIONcaptionsoff
  \newpage
\fi

\bibliographystyle{./bibliography/IEEEtran}
\bibliography{./bibliography/Dynamic_Antenna.bib,./bibliography/IEEEabrv.bib,./bibliography/IEEEexample.bib}

\end{document}